\documentclass[aip,graphicx,reprint,jcp]{revtex4-1}

\usepackage{graphicx}
\usepackage{amsmath}
\usepackage{amssymb}
\usepackage{chemarr}
\usepackage[english]{babel}
\usepackage{color}

\draft % marks overfull lines with a black rule on the right

\begin{document}

%\preprint{}

\title{Binding constants of membrane-anchored receptors and ligands: a general theory corroborated by Monte Carlo simulations} %Title of paper

\author{Guang-Kui Xu}
\affiliation{Max Planck Institute of Colloids and Interfaces, Department of Theory and Bio-Systems, 14424 Potsdam, Germany}
\affiliation{International Center for Applied Mechanics, State Key Laboratory for Strength and Vibration of Mechanical Structures, Xi'an Jiaotong University, Xi'an 710049, China}
\author{Jinglei Hu}
\affiliation{Max Planck Institute of Colloids and Interfaces, Department of Theory and Bio-Systems, 14424 Potsdam, Germany}
\affiliation{Kuang Yaming Honors School, Nanjing University, 210023 Nanjing, China}
\author{Reinhard Lipowsky}
\affiliation{Max Planck Institute of Colloids and Interfaces, Department of Theory and Bio-Systems, 14424 Potsdam, Germany}
\author{Thomas R.\ Weikl}
\affiliation{Max Planck Institute of Colloids and Interfaces, Department of Theory and Bio-Systems, 14424 Potsdam, Germany}

%\date{\today}

\begin{abstract}
Adhesion processes of biological membranes that enclose cells and cellular organelles are essential for immune responses, tissue formation, and signaling. These processes depend sensitively on the binding constant $K_\text{2D}$ of the membrane-anchored receptor and ligand proteins that mediate adhesion, which is difficult to measure in the `two-dimensional' (2D) membrane environment of the proteins. An important problem therefore is to relate $K_\text{2D}$ to the binding constant $K_\text{3D}$ of soluble variants of the receptors and ligands that lack the membrane anchors and are free to diffuse in three dimensions (3D). In this article, we present a general theory for the binding constants $K_\text{2D}$ and $K_\text{3D}$ of rather stiff proteins whose main degrees of freedom are translation and rotation, along membranes and around anchor points `in 2D', or unconstrained `in 3D'. The theory generalizes previous results by describing how $K_\text{2D}$ depends both on the average separation and thermal nanoscale roughness of the apposing membranes, and on the length and anchoring flexibility of the receptors and ligands. Our theoretical results for the ratio  $K_\text{2D}/K_\text{3D}$ of the binding constants agree with detailed results from Monte Carlo simulations without any data fitting, which indicates that the theory captures the essential features of the `dimensionality reduction' due to membrane anchoring. In our Monte Carlo simulations, we consider a novel coarse-grained model of biomembrane adhesion in which the membranes are represented as discretized elastic surfaces, and the receptors and ligands as anchored molecules that diffuse continuously along the membranes and rotate at their anchor points. 
\end{abstract}


\maketitle %\maketitle must follow title, authors, abstract and \pacs


\section{Introduction}

Cell adhesion and the adhesion of vesicles to the membranes of cells and cellular organelles is mediated by the binding of receptor and ligand proteins that are anchored in the adhering membranes. Central questions are how the binding affinity of the anchored proteins can be measured and quantified, how this affinity is affected by characteristic properties of the proteins and membranes, and how it is related to the affinity of soluble variants of the receptor and ligand proteins without membrane anchors \cite{Dustin01,Orsello01,Leckband12,Zarnitsyna12,Krobath09,Wu10,Hu13,Wu13}. For soluble receptors and ligands that are free to diffuse in three dimensions (3D), the binding affinity can be quantified by the binding equilibrium constant 
\begin{equation}
 K_\text{3D} = \frac{[\text{RL}]_\text{3D}}{[\text{R}]_\text{3D}[\text{L}]_\text{3D}}
\label{K3D}
\end{equation}
where $[\text{RL}]_\text{3D}$ is the {\em volume} concentration of bound receptor-ligand complexes, and $[\text{R}]_\text{3D}$ and $[\text{L}]_\text{3D}$ are the volume concentrations of unbound receptors and unbound ligands in the solution. The binding constant $K_\text{3D}$ is determined by the binding free energy of the complex and, thus, by local interactions at the binding sites of the proteins, at least in the absence of more global conformational changes of the proteins during binding. The binding constant $K_\text{3D}$ can be measured with standard experimental methods \cite{Schuck97,Rich00,McDonnell01}.  A two-dimensional (2D) analogue for membrane-anchored receptors and ligands that are restricted to the membrane environment is the binding constant
\begin{equation}
K_\text{2D} = \frac{[\text{RL}]_\text{2D}}{[\text{R}]_\text{2D}[\text{L}]_\text{2D}}
\label{K2Ddef}
\end{equation}
where $[\text{RL}]_\text{2D}$,  $[\text{R}]_\text{2D}$, and $[\text{L}]_\text{2D}$ are the {\em area} concentrations of bound receptor-ligand complexes, unbound receptors, and unbound ligands \cite{Dustin01,Orsello01}. The binding of membrane-anchored receptors and ligands in cell adhesion zones has been experimentally investigated with fluorescence methods \cite{Dustin96,Dustin97,Zhu07,Tolentino08,Huppa10,Axmann12,ODonoghue13} and with several mechanical methods involving hydrodynamic flow  \cite{Kaplanski93,Alon95}, centrifugation \cite{Piper98}, or micropipette setups that use red blood cells as force sensors \cite{Chesla98,Merkel99,Williams01,Chen08,Huang10,Liu14}. However, the $K_\text{2D}$ values obtained from different methods can differ by several orders of magnitude\cite{Dustin01}, which indicates a `global'  dependence of $K_\text{2D}$ on the membrane adhesion system, besides the dependence on local receptor and ligand interactions. 

In this article, we present a general theory that relates the binding constant $K_\text{2D}$ of membrane-anchored receptor and ligand molecules to the binding constant $K_\text{3D}$ of soluble variants of these molecules. This theory describes how $K_\text{2D}$ depends both on overall characteristics of the membranes and on molecular properties of the receptors and ligands. Quantifying $K_\text{2D}$ is complicated by the fact that the binding of membrane-anchored receptors and ligands depends on the local separation $l$ of the membranes, which varies -- along the membranes, and in time -- because of thermally excited membrane shape fluctuations. Experiments that probe $K_\text{2D}$ imply averages in space and time over membrane adhesion regions and measurement durations. In our theory, we first determine the binding constant $K_\text{2D}$ for a given local separation $l$, and then average over the distribution $P(l)$ of local membrane separations that describes the spatial and temporal variations of $l$. The two key overall membrane characteristics that emerge from this theoretical approach are the average separation $\bar{l}$ and relative roughness $\xi_\perp$ of the two apposing membranes, which are the mean and standard deviation of the distribution $P(l)$. Our theory quantifies the dependence of $K_\text{2D}$ on the average separation $\bar{l}$ and relative membrane roughness $\xi_\perp$, and helps to understand why different experimental methods can lead to values of $K_\text{2D}$ that differ by orders of magnitude \cite{Dustin01} { (see Discussion and Conclusions)}. 

Our theory is validated in this article by a detailed comparison to data from Monte Carlo (MC) simulations. Such a comparison is essential to test simplifying assumptions and heuristic elements in relating $K_\text{2D}$ to the binding constant $K_\text{3D}$ of soluble variants of receptors and ligands without membrane anchors. Our theoretical results for the ratio  $K_\text{2D}/K_\text{3D}$ of the binding constants agree with detailed results from MC simulations without any data fitting, which indicates that our theory captures the essential features of the `dimensionality reduction' due to membrane anchoring. The MC simulations are based on a novel model of biomembrane adhesion in which the membranes are represented as discretized elastic surfaces, and the receptors and ligands as anchored molecules that diffuse continuously along the membranes and rotate around their anchoring points. We use the MC simulations to determine both the binding constant $K_\text{2D}$ of these membrane-anchored molecules and the binding constant $K_\text{3D}$ of soluble variants of the molecules that have the same binding interactions but are free to move in 3D. In previous elastic-membrane models of biomembrane adhesion, determining both $K_\text{2D}$ and $K_\text{3D}$ and the molecular characteristics affecting these binding constants has not been possible because the receptors and ligands are not explicitly represented as anchored molecules. Instead, the binding of receptors and ligands has been described implicitly by interactions that depend on the membrane separation \cite{Lipowsky96,Weikl01,Weikl02a,Weikl04,Asfaw06,Tsourkas07,Reister08,Bihr12}. In other previous elastic-membrane models, receptors and ligands are described by concentration fields rather than individual molecules \cite{Komura00,Bruinsma00,Qi01,Chen03,Raychaudhuri03,Coombs04,Shenoy05,Wu06}, or receptor-ligand bonds are treated as constraints on the local membrane separation \cite{Zuckerman95,Krobath07}.
{ In our accompanying article\cite{Hu}, we compare our theory for the binding equilibrium of membrane-anchored receptor and ligand molecules to detailed data from molecular dynamics simulations of a coarse-grained molecular model of biomembrane adhesion \cite{Hu13}, and extend this theory to the binding kinetics of membrane-anchored molecules.}

\begin{figure*}[t]
\resizebox{1.6\columnwidth}{!}{\includegraphics{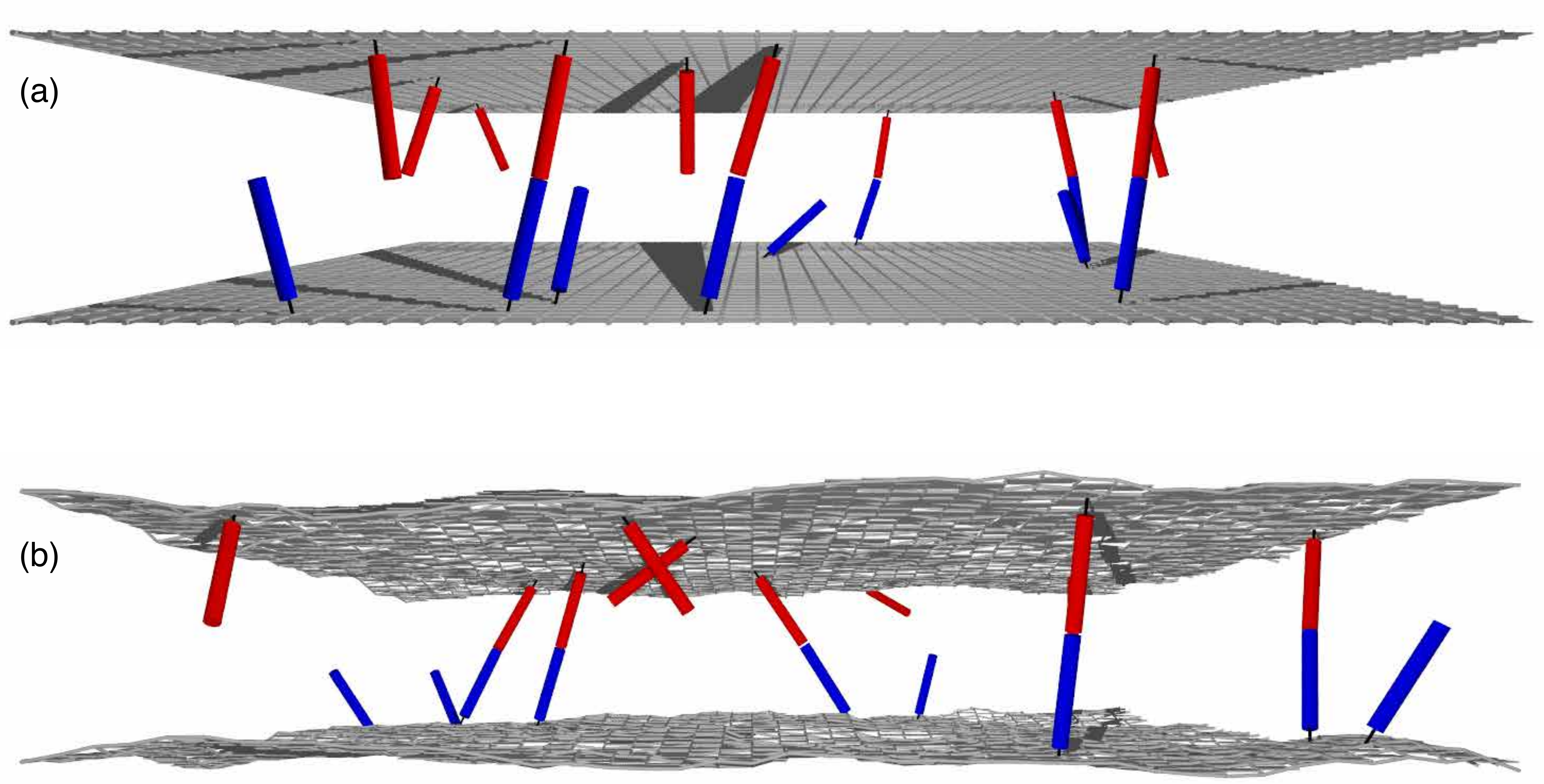}}
%\resizebox{\columnwidth}{!}{\includegraphics{figure1}}
\caption{Snapshots from MC simulations with rigid receptors and ligands anchored (a) to parallel and planar membranes with separation $l= 7.8\,a$, and (b) to fluctuating membranes. The length of the receptors and ligands is $L_R = L_L = 4\,a$, and the anchoring strength is $k_a = 4\,k_B T$. The snapshots display membrane segments of area $40 \times 40\,a^2$ from simulations with overall membrane area $160 \times 160\,a^2$ and 200 receptors and ligands. }
\label{figure_MCsnapshots}
\end{figure*}
%

%%%
\section{Coarse-grained elastic-membrane model of biomembrane adhesion}
%%%

In this section, we introduce our elastic-membrane model of biomembrane adhesion. In this model, the overall configurational energy of rod-like receptors and ligands
\begin{equation}
H=H_\text{el}^{(1)} +H_\text{el}^{(2)} +H_\text{int}+H_\text{anc}
\label{overall_energy}
\end{equation}
is the sum of the elastic energies $H_\text{el}^{(1)}$ and $H_\text{el}^{(2)}$ of the two membranes, the total interaction energy $H_\text{int}$ of the receptor and ligand molecules, and the total anchoring energy $H_\text{anc}$ of these molecules. 

\subsection{Elastic energy of the membranes.} 

The conformations of the two apposing membranes can be described in Monge representation via their local deviations out of a reference plane. We discretize this reference plane into a quadratic lattice with lattice spacing $a$, which results in a partitioning of the membranes into approximately quadratic patches. The elastic energy  $H_\text{el}^{(1)}$ and $H_\text{el}^{(2)}$ of the membranes then can be written as \cite{Lipowsky89,Weikl06}
\begin{equation}
H_\text{el}^{(j)} = \sum_i\left[(\kappa^{(j)}/2a^2)(\Delta_d l_i^{(j)})^2 + (\sigma^{(j)}/2)(\nabla_d l_i^{(j)})^2\right] 
\label{elastic_energy}
\end{equation}
with $j=1, 2$ where $l_i^{(1)}$ and $l_i^{(2)}$ are the local deviations of the membranes at lattice site $i$ out of the reference plane. The elastic energy (\ref{elastic_energy}) is the sum of the bending energy with rigidity $\kappa^{(j)}$ \cite{Helfrich73} and the contribution from the membrane tension $\sigma^{(j)}$. The bending energy depends on the total curvature $\Delta_d l_i/a^2$ with discretized Laplacian
\begin{equation}
\Delta_d l_i=\Delta_d l_{x,y} =l_{x+a,y}+l_{x-a,y}+l_{x,y+a}+l_{x,y-a}-4l_{x,y}
\end{equation}
The tension contribution depends on the local area increase
\begin{equation}
(\nabla_d l_i)^2=(\nabla_d l_{x,y})^2=(l_{x+a,y}-l_{x,y})^2 + (l_{x,y+a}-l_{x,y})^2
\end{equation}
of the curved membranes with respect to the reference $x$-$y$ plane. The whole spectrum of bending deformations is captured in this model if the lattice spacing $a$ of the discretized membranes is about 5 nm, which is close to the membrane thickness \cite{Goetz99}.

\subsection{Binding and anchoring of receptors and ligands.}

The total interaction energy $H_\text{int}$ represents the interactions of all receptor-ligand complexes. In our model, the binding potential of a single receptor and a single ligand 
\begin{equation}
V_\text{int}(r,\theta_1,\theta_2)=U_b e^{-k_r r^2/2} e^{-k_{\theta}(\theta_1^2+\theta_2^2)/2}
\label{binding_potential}
\end{equation}
depends on the distance $r$ between the binding sites located at the tips of the rod-like receptor and ligand molecules, and on the two angles $\theta_1$ and $\theta_2$ that describe the relative orientation of the molecules. For our rod-like receptors and ligands, the angle $\theta_1$ is the angle between the receptor and the binding vector connecting the two binding sites, and the angle $\theta_2$ is the angle between the ligand and this vector. We use two angles $\theta_1$ and $\theta_2$ for the relative orientation to ensure that the binding sites of the receptor and ligand do not overlap. The total interaction energy $H_\text{int}$ of the receptors and ligands in Eq.\ (\ref{overall_energy}) is the sum of the potential energies (\ref{binding_potential}) of all bound receptor-ligand complexes. 

The total anchoring energy $H_\text{anc}$ is the sum of the anchoring energies of all receptors and ligands. In our model, the anchoring energy of a single receptor or ligand is described by the harmonic potential 
\begin{equation}
V_\text{anchor}=\frac {1}{2} k_a \theta_a^2
\label{Vanchor}
\end{equation}
with anchoring strength $k_a$. The anchoring angle $\theta_a$ is the angle between the receptors or ligands and the local membrane normal (see 
Appendix A for further details).

%%%
\section{General theory for the binding constants of rigid receptors and ligands}
%%%

In this section, we derive our general theory for the binding constants $K_\text{2D}$ and $K_\text{3D}$ of rigid, rod-like receptors and ligands. 
The starting point of our theory is the binding free energy  $\Delta G_{\rm 2D}$ and $\Delta G_{\rm 3D}$ of membrane-anchored and soluble receptor and ligand molecules. We first summarize a standard theory for the binding free energy $\Delta G_{\rm 3D}$ of soluble molecules, and then extend this theory to the binding free energy $\Delta G_{\rm 2D}$ of membrane-anchored molecules. From these binding free energies, we obtain general relations between the binding constants $K_\text{2D}$ and $K_\text{3D}$. In section IV, we compare these theoretical relations to detailed results from MC simulations, and generalize our theory to semi-flexible receptor and ligand molecules. 

\subsection{Binding free energy of soluble receptors and ligands.} 

We first consider the binding free energy $\Delta G_{\rm 3D}$ of a single soluble receptor and a single soluble ligand in a volume $V$. A standard approach in which this free energy is expanded around its minimum leads to the decomposition \cite{Hu13,Luo02, Woo05}
\begin{align}
\Delta G_{\rm 3D} &\simeq U_0 + \Delta G_{\rm trans} + \Delta G_{\rm rot} \nonumber \\
&\simeq U_0 - k_B T \ln\left[\frac{V_b}{V}\right] - k_BT \ln\left[\frac{\Omega_b}{4 \pi}\right]
\label{DG3D}
\end{align}
into the minimum binding energy $U_0$ and the translational and rotational free-energy contributions $\Delta G_{\rm trans}$ and $\Delta G_{\rm rot}$. Here, $V_b$ and $\Omega_b$ are the translational and rotational phase-space volume of the bound ligand relative to the receptor. The translational phase-space volume of the bound ligand is $V_b = (2 \pi)^{3/2} \xi_x \xi_y \xi_z$ where $\xi_x$, $\xi_y$, and $\xi_z$ are the standard deviations of the distributions for the coordinates $x$, $y$, and $z$ of the binding vector that connects the two binding sites. The $z$-direction here is taken to be parallel to the direction of the receptor-ligand complex. For a preferred collinear binding of the receptor and ligand as in the binding potential of Eq.\ (\ref{binding_potential}), the rotational phase space volume of the bound ligand is $\Omega_b = 2 \pi \sigma^2_b$ where  $\sigma_b$ is the standard deviation of the binding-angle distribution \cite{Hu13}. The unbound ligand translates and rotates freely with translational phase-space volume $V$ and rotational phase-space volume $4\pi$. 

\subsection{Binding free energy of receptors and ligands anchored to planar and parallel membranes.}

In analogy to Eq.\ (\ref{DG3D}), the binding free energy $\Delta G_{\rm 2D}$ of a receptor and a ligand molecule that are anchored to two apposing planar and parallel membranes of area $A$ and separation $l$ can be decomposed as \cite{Hu13}
\begin{align}
\Delta G_{\rm 2D} &\simeq U_0 + \Delta G_{\rm trans} + \Delta G_{\rm rot} \nonumber \\
&\simeq U_0 - k_B T \ln\left[\frac{A_b}{A}\right] - k_B T\ln\left[\frac{\Omega_b\Omega_\text{RL}}{\Omega_\text{R}\Omega_\text{L}}\right]
\label{DG2D}
\end{align}
where $A_b = 2 \pi \xi_x \xi_y$ is the translational phase space area of the bound ligand relative to the receptor in the two
directions $x$ and $y$ parallel to the membranes, and $\Omega_\text{R}$, $\Omega_\text{L}$, and $\Omega_\text{RL}$ are the rotational phase space volumes of the unbound receptor R, unbound ligand L, and bound receptor-ligand complex RL relative to the membranes. We have assumed here that the binding angle variations are small compared to the overall rotations of the bound RL complex, i.e.~we have assumed that the anchoring potential is `soft' compared to the binding potential. The rational phase space volume $\Omega_b$ for the binding angle and the minimal binding energy $U_0$ then are not affected by the anchoring, and the overall rotational phase space volume of the bound complex can be approximated as the product of the rotational phase space volume $\Omega_b$ for the binding angle and the phase space volume $\Omega_\text{RL}$ for the rotations of the whole complex relative to the membrane \cite{Hu13}. For the harmonic anchoring potential ({\ref{Vanchor}}), the rotational phase space volumes of the unbound molecules  are
\begin{align}
\Omega_\text{R} = \Omega_\text{L} &=  2\pi \int_0^{\pi/2} e^{-\frac{1}{2} k_a \theta_a^2/k_B T} \sin \theta_a \, \text{d}\theta_a  \label{OmegaR}
\\
& \simeq 2\pi k_B T/k_a
\text{~~for~~} k_a \gg k_B T 
\end{align}
For simplicity, we consider here receptors and ligands with identical anchoring strength $k_a$.

The remaining task now is to determine the phase space volume $\Omega_\text{RL}$ for the rotations of the bound RL complex relative to the membrane. We find that these rotations can be described by the effective configurational energy (see Appendix B)
\begin{equation}
H_\text{RL}(\theta_a,L_\text{RL}) \simeq k_a \theta_a^2 + \frac{1}{2} k_\text{RL}(L_\text{RL} - L_0)^2
\label{Hef}
\end{equation}
The first term of this effective energy is the sum of the anchoring energies ({\ref{Vanchor}}) for the receptor and ligand in the complex. The two anchoring angles $\theta_a$ for the bound receptor and ligand here are taken to be approximately equal, which holds for binding angles and binding  angle variations that are small compared to the anchoring angle variations, or in other words,  for binding potentials that are `hard' compared to the anchoring potentials. The second term of the effective energy (\ref{Hef}) is a harmonic approximation for variations in the length $L_\text{RL}$ of the receptor-ligand complex, i.e.~in the distance between the two anchoring points of the complex. For rod-like receptor and ligand molecules, variations in the length $L_\text{RL}$ of the complex result from variations of the binding angle and binding-site distance. The preferred length $L_0$ and effective spring constant $k_\text{RL}$ of the RL complex in the effective energy (\ref{Hef}) are then approximately (see Appendix B)
\begin{align}
 L_0  \simeq L_\text{R} + L_\text{L} + z_0 - \sigma_b^2 L_\text{R} L_\text{L}/(L_\text{R} + L_\text{L}) 
 \label{L0}\\
 k_\text{RL}  \simeq  k_B T/\left(\xi_z^2 + \sigma_b^4 L_\text{R}^2 L_\text{L}^2/(L_\text{R} + L_\text{L})^2\right)
 \label{kRL}
\end{align}
where $L_\text{R}$ and $L_\text{L}$ are the lengths of the rod-like receptor and ligand, $z_0$ is the average of the distance between the binding sites in the direction of the complex, $\xi_z$ is the standard deviation of this distance, and $\sigma_b$ is the standard deviation of the binding-angle distribution for preferred collinear binding as in our model. 

For a given separation $l$ of the membranes, the length $L_\text{RL}$ and anchoring angle $\theta_a$ of the receptor-ligand complex are related via
\begin{equation}
L_\text{RL}(\theta_a) = l /\cos \theta_a
\end{equation}
The effective configurational energy (\ref{Hef}) then only depends on the single variable $\theta_a$. With this effective configurational energy, the rotational phase space volume of the bound RL complex can be calculated as
\begin{equation}
\Omega_\text{RL}(l) \simeq 2 \pi \int_0^{\pi/2} e^{-H_\text{RL}(\theta_a,L_\text{RL}(\theta_a))/k_B T}  \sin\theta_a \text{d}\theta_a
\label{OmegaRL}
\end{equation}
The integration in Eq.~(\ref{OmegaRL}) can be easily evaluated numerically for specific values of the spring constants $k_a$ and $k_\text{RL}$, of the preferred length $L_0$ of the complex, and of the membrane separation $l$. 

\subsection{Binding constant of receptors and ligands anchored to planar and parallel membranes.}

From the binding free energies $\Delta G_{\rm 2D}$ and $\Delta G_{\rm 3D}$ given in Eqs.\ (\ref{DG3D}) and (\ref{DG2D})  and the relations  $K_\text{2D} = A \exp[-\Delta G_{\rm 2D} /k_B T]$ and $K_\text{3D} = V \exp[-\Delta G_{\rm 3D} /k_B T]$ between the binding free energies and binding constants \cite{Hu13}, we obtain the general result
\begin{equation}
K_\text{2D}(l) \simeq K_\text{3D}\frac{\sqrt{8 \pi}}{\xi_z}\frac{\Omega_\text{RL}(l)}{\Omega_\text{R}\Omega_\text{L}}
\label{K2Dl}
\end{equation}
which relates the binding constant $K_\text{2D}(l)$ of receptors and ligands anchored to parallel and planar membranes of separation $l$ to the binding constant $K_\text{3D}$ of soluble variants of the receptors and ligands without membrane anchors. In deriving Eq.\ (\ref{K2Dl}), we have assumed that the binding interface is not affected by the membrane anchoring, which holds for anchoring potentials that are much softer than the binding potential. The minimum binding energy $U_0$ and the standard deviations $\xi_x$ and $\xi_y$ of the binding vector coordinates in the two directions perpendicular to the complex are then the same for the soluble and the membrane-anchored receptor-ligand complex. For simplicity, we take the two directions $x$ and $y$  perpendicular to the complex to be identical with the two directions along the membranes. The ratio of the translational phase space volume of the soluble RL complex and the translational phase space area of the bound complex  then is approximately $V_b/A_b \simeq \sqrt{2\pi} \xi_z$.\footnote{The effect of the tilt of the receptor-ligand complexes relative to the membrane normal on $V_b/A_b$ can be taken into account via $\xi_x$ and $\xi_y$. However, since the values of the standard deviations $\xi_x$, $\xi_y$, and $\xi_z$ in the directions $x$ and $y$ perpendicular to the complex and the direction $z$ parallel to the complex are typically rather similar, we neglect this effect here.} 

\subsection{Binding constant of receptors and ligands anchored to fluctuating membranes.}

In membrane-membrane adhesion zones, the local separation $l$ is not  fixed but varies because of thermally excited shape fluctuations of the membranes. { Our MC simulations show that} the distribution $P(l)$ of this local  separation is well approximated by the Gaussian distribution
\begin{equation}
P(l) \simeq \exp\left[-(l-\bar l)^2/2\xi_\perp^2\right]/(\sqrt{2\pi} \xi_\perp)
\label{pl}
\end{equation}
where $\bar{l}=\langle l \rangle$ is the average separation of the membranes or membrane segments, and $\xi_\perp = \sqrt{\langle(l - \bar{l})^2\rangle}$ is the relative roughness of the membranes. The relative roughness is the standard deviation of the local membrane separation $l$, i.e.~the width of the distribution $P(l)$. { The same Gaussian behavior of $P(l)$ is also found in molecular dynamics simulations (see our accompanying manuscript \cite{Hu}).} The Gaussian behavior of $P(l)$ holds for situations in which the adhesion of two apposing membrane segments is mediated by a single type of receptors and ligands as in our simulations.

{ Our MC simulations also reveal that the equilibrium constant $K_\text{2D}$ for fluctuating membranes can be obtained in two rather different ways. On the one hand, we can determine $K_\text{2D}$ directly from its definition in Eq. (\ref{K2Ddef}) by measuring the area concentrations  $[\text{RL}]$, $[\text{R}]$, and $[\text{L}]$ in the simulations. On the other hand, this equilibrium constant can also be obtained from the equilibrium constants $K_\text{2D}(l)$ for planar membranes  via the simple relation} 
\begin{equation}
K_\text{2D} = \int K_\text{2D}(l) P(l) \text{d}l
\label{K2D}
\end{equation}
{ i.e., by averaging $K_\text{2D}(l)$ over the distribution $P(l)$  for the local membrane separation. The relation in Eq. (\ref{K2D}) implies that we can identify the constant separation of the planar membranes with the local separation of the fluctuating membranes. This conclusion is somewhat surprising because thermally excited shape fluctuations of the membranes also lead to fluctuations of the membranes' normal vectors, which { affect} the energetically most favorable local orientations of the receptor and ligand molecules. However, as shown in the Appendices D and E, the contribution from these orientational fluctuations is relatively small and can be neglected compared to the fluctuations in the local separation $l$.}  
If we ignore the orientational fluctuations of the membranes, Eq.\ (\ref{K2D}) can also be justified by the fact that the calculation of thermodynamic equilibrium quantities such as $K_\text{2D}$ does not depend on the order in which the degrees of freedom of a system are averaged \footnote{In contrast, related averages over local membrane separations for the on-rate constant $k_\text{on}$ and off-rate constant $k_\text{off}$ rely on characteristic timescales for membrane shape fluctuations that are much smaller than the characteristic timescales for the diffusion of the anchored molecules on the relevant length scales, and much smaller than the characteristic binding times \cite{Bihr12,Hu}}.
Eq.\ (\ref{K2D}) implies that the translational and rotational degrees of freedom of the receptors and ligands are averaged first to calculate $K_\text{2D}(l)$ given in Eq.\ (\ref{K2Dl}), followed by a second average over the local membrane separations $l$ with probability distribution $P(l)$. We thus propose that Eq.\ (\ref{K2D}) is general and holds for any shape of the distribution $P(l)$.

For a relative membrane roughness $\xi_\perp$ that is much larger than the width $\xi_\text{RL}$ of the function $K_\text{2D}(l)$, the distribution $P(l)$ is nearly constant over the range of local separations $l$ for which $K_\text{2D}(l)$ is not negligibly small. The average over local separations in Eq.\ (\ref{K2D}) for the Gaussian distribution (\ref{pl}) of $P(l)$
then simplifies to (see Appendix C)
\begin{equation}
K_\text{2D} \simeq P(\bar{l}_0) \int K_\text{2D}(l) \text{d}l \simeq \frac{K_\text{3D} k_a }{\sqrt{2\pi k_B T k_\text{RL}}\, \xi_z\xi_\perp} e
^{-(\bar{l}_0 -\bar l)^2/2\xi_\perp^2}
\label{K2Dlim}
\end{equation}
for anchoring strengths $k_a \gg k_B T$, where $\bar{l}_0$ is the preferred average separation of the receptor-ligand complexes for large membrane roughnesses. For such large roughnesses $\xi_\perp$ and anchoring strengths $k_a$, the preferred average separation of the receptor-ligand complexes is (see Appendix C)
\begin{equation}
\bar{l}_0 \simeq L_0(1-k_BT/2 k_a)
\label{barl0}
\end{equation}
This preferred average separation is smaller than the preferred length $L_0$ of the receptor-ligand complexes because of the tilting of the complexes.
The width of the function $K_\text{2D}(l)$ can be estimated as the standard deviation (see Appendix C)
\begin{equation}
\xi_\text{RL}\simeq \sqrt{(k_B T/k_\text{RL}) + (k_BT L_0/2k_a)^2}
\label{xiRL}
\end{equation}
for large anchoring strengths $k_a \gg k_B T$.

\section{MC data for the binding constants of membrane-anchored receptors and ligands}

In this section, we compare our theoretical results to detailed data from MC simulations with membrane-anchored receptor and ligand molecules. These data result from two different simulation scenarios: First, we have performed MC simulations with two apposing parallel and planar membranes to determine the binding constant $K_\text{2D}$ as a function of the local membrane separation $l$ (see Fig.\ \ref{figure_MCsnapshots}(a)). In these simulations, the local separation $l$ is constant for all membrane patches and, thus, identical to the average separation $\bar{l}$ of the membranes. By varying the membrane separation $l$, we obtain the function $K_\text{2D}(l)$ from these simulations. Second, we have performed MC simulations with flexible membranes in which the local separation $l$ of the apposing membranes varies because of thermally excited shape fluctuations of the membranes (see Fig.\ \ref{figure_MCsnapshots}(b)). These variations can be quantified by the relative roughness $\xi_\perp$ of the membranes, which is the standard deviation of the local separation. The relative roughness in our simulations depends on the number of bound receptor-ligand complexes, because the complexes constrain the shape fluctuations, and on the membrane tension, which suppresses such fluctuations. In these simulations, the membranes are `free to choose'  an optimal average separation $\bar{l}_0$ at which the overall free energy is minimal. We thus obtain $K_\text{2D}$ as a function of the membrane roughness $\xi_\perp$ at the average membrane separation $\bar{l}=\bar{l}_0$ from these simulations. 

We use the parameter values  $U_b = - 25$ $k_B T$, $k_r = 20$ $k_BT/a^2$, and $k_{\theta}=15 k_B T$ for the binding potential (\ref{binding_potential}) of receptors and ligands in all our simulations. For these parameter values, the average distance between the two binding sites in the direction of the receptor-ligand complex is $z_o = 0.078$ $a$, the standard deviation of this average distance is $\xi_z = 0.036$ $a$, and the standard deviation of the binding angle, i.e.~the angle between the receptor and ligand at the interaction sites, is $\sigma_b = 0.080$. The binding potential  (\ref{binding_potential}) with these parameter values is rather `hard' compared to the anchoring potential (\ref{Vanchor}) with anchoring strengths $k_a= 4\, k_BT$, $8\, k_BT$, or $16\, k_BT$ considered in our simulations. The average distance $z_o$ of the  binding sites in the direction of the complex and the standard deviations $\xi_z$ and $\sigma_b$ then do not depend on whether the receptor and ligand molecules are anchored to membranes, or soluble. The direction of the receptor-ligand complex is the direction of the line connecting the two anchoring sites at the ends of the complex. { The values of the anchoring strength $k_a$ considered in our simulations are within the  range of anchoring strengths obtained from coarse-grained molecular dynamics simulations with lipid-anchored and transmembrane receptors and ligands \cite{Hu}.}

We determine the binding constant $K_\text{2D}$ of the membrane-anchored receptors and ligands with Eq.\ (\ref{K2Ddef}). The area concentrations  $[\text{RL}]_\text{2D}$,  $[\text{R}]_\text{2D}$,  and $[\text{L}]_\text{2D}$ in this equation are obtained from thermodynamic averages of the numbers of receptor-ligand complexes, of unbound receptors, and of unbound ligands for the membrane area $160 \times 160$ $a^2$ of our simulations with periodic boundary conditions. We define a receptor and ligand to be bound if the binding distance $r$ and the two angles $\theta_1$ and $\theta_2$ in Eq.\ (\ref{binding_potential}) are smaller than the cutoff values $r_c=0.58$ $a$ and $\theta_c=0.67$, respectively. These cutoff values include 99\% of the area of the Gaussian functions $\exp(-k_r r^2/2)$ and $\exp(-k_{\theta}\theta_i^2/2)$ in Eq.\ (\ref{binding_potential}) for the parameter values $k_r = 20$ $k_BT/a^2$ and $k_{\theta}=15 k_B T$ used in our simulations. We only allow the binding of a single ligand to a single receptor. In our simulations, the numbers $N_R$ and $N_L$ of receptors and ligands varies between $N_R = N_L = 50$ and $1000$. The binding constant $K_\text{3D}$ of soluble variants of the receptors and ligands is determined from Eq.\ (\ref{K3D}). These soluble receptor and ligand molecules exhibit the same binding potential (\ref{binding_potential}) as the membrane-anchored molecules, but translate and rotate freely in a box of volume $V$ with periodic boundary conditions. For the parameters of the binding potential given above, we obtain the value $K_\text{3D} \simeq 261$ $a^3$. 

For simplicity, we consider two membranes with identical rigidity $\kappa$ and tension $\sigma$ in our simulations with flexible membranes. We use the value $\kappa = 10$ $k_BT$ in all our simulations, which is a typical value for lipid membranes \cite{Dimova14}. Our MC simulations with flexible membranes involve three types of MC moves: (i) The lateral diffusion of a  receptor or ligand along the membranes is taken into account by moves in which the coordinates $(x_a, y_a)$ of the anchoring points in the reference plane are continuously and randomly shifted to new values. The local deviation $l_a$ of this anchoring point in the direction perpendicular to the reference plane is determined by linear interpolation of the local deviations $l_i$ of the discretized membranes (see Appendix A for further details). (ii) The rotational diffusion of the rod-like receptors and ligands is taken into account by random continuous rotational moves around the anchor points. (iii) Shape fluctuations of the membranes can be taken into account by moves in which the local deviations $l_i^{(1)}$ and $l_i^{(2)}$ are randomly shifted to new values \cite{Weikl06}. Our MC simulations with parallel and planar membranes only involve the MC moves (i) and (ii).

\begin{figure}[t]
\resizebox{\columnwidth}{!}{\includegraphics{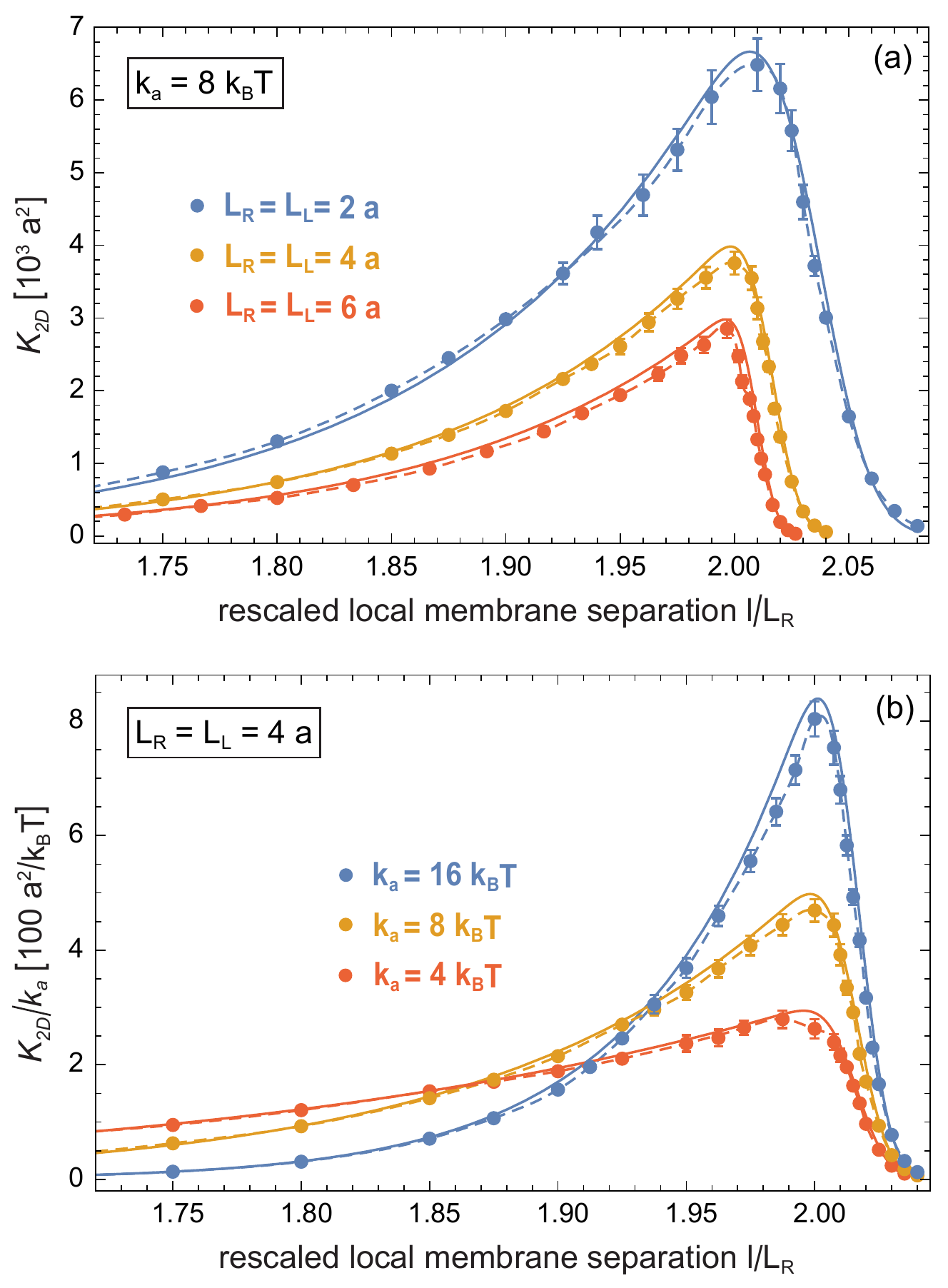}}
%\resizebox{0.6\columnwidth}{!}{\includegraphics{figure2}}
\caption{Binding constant $K_\text{2D}$ as a function of the local membrane separation $l$  for (a) the anchoring strength $k_a = 8 \, k_B T$ and different lengths $L_R = L_L$ of rigid receptors and ligands, and for (b) $L_R = L_L= 4 \, a$ and different values of $k_a$. The local separation $l$ is rescaled by the length $L_R = L_L$ of receptors and ligands. In (b), the binding constant $K_\text{2D}$ is rescaled by $k_a$. The full lines result from Eq.\ (\ref{K2Dl}) of our theory. The dashed lines are interpolations of the MC data points. { The MC data points are obtained from simulations with parallel and planar membranes (see Fig.\ \ref{figure_MCsnapshots}(a)) based on Eq.\ (\ref{K2Ddef}).}
}
\label{figure_MC-planar}
\end{figure}
\subsection{Binding constant of rigid receptors and ligands as a function of the local membrane separation. }  

We first consider results from our MC simulations with rigid, rod-like receptors and ligands anchored to parallel and planar membranes. In Fig.\ \ref{figure_MC-planar}, MC data for the function $K_\text{2D}(l)$ are compared to our theory for various values of the anchoring strength $k_a$ and length $L_\text{R} = L_\text{L}$ of the receptors and ligands. The full lines in this figure result from Eq.\ (\ref{K2Dl}) of our theory and do not involve any fit parameters. The dashed lines in the figure are interpolations of the MC data points. For the binding potential of the receptors and ligands used in our simulations, the average distance between the two binding sites in the direction of the receptor-ligand complex is $z_o = 0.078$ $a$, the standard deviation of this distance is $\xi_z = 0.036$ $a$, the standard deviation of the binding angle is $\sigma_b = 0.080$, and the binding constant of soluble variants of the receptors and ligands is $K_\text{3D}\simeq 261$ $a^3$ (see above). With these values for $z_o$, $\xi_z$,  $\sigma_b$, and $K_\text{3D}$, the function $K_\text{2D}(l)$ can be calculated from the Eqs.\ (\ref{OmegaR}), (\ref{L0}), (\ref{kRL}), (\ref{OmegaRL}), and (\ref{K2Dl}) of our theory for the  various anchoring strengths $k_a$ and molecular lengths $L_\text{R} = L_\text{L}$ of Fig.\ \ref{figure_MC-planar}. The function $K_\text{2D}(l)$ exhibits a maximum value at a preferred local separation $l_0$ of the receptors and ligands, and is asymmetric with respect to $l_0$. This asymmetry reflects that the receptor-ligand complexes can tilt at local separations $l$ smaller than $l_0$, but need to stretch at local separations larger than $l_0$. 

Fig.\ \ref{figure_MC-planar} illustrates that $K_\text{2D}(l)$ strongly depends both on the length $L_\text{R} = L_\text{L}$ and anchoring strength $k_a$ of the receptors and ligands. The decrease of $K_\text{2D}(l)$ for increasing length $L_\text{R} = L_\text{L}$ results from a decrease of the rotational phase space volume $\Omega_\text{RL}(l)$ of the receptor-ligand complex. With increasing length of the receptors and ligands, the RL complexes become effectively stiffer because $k_\text{RL} L_0^2$ in Eq.\ (\ref{OmegaRL}) increases from $12.5 \cdot 10^3\,k_BT$ for $L_\text{R} = L_\text{L}=2\ a$ to $44.9 \cdot 10^3\,k_BT$ and $87.8 \cdot 10^3\,k_BT$ for $L_\text{R} = L_\text{L}=4\,a$ and $6\,a$, respectively. The effective stiffness $k_\text{RL} L_0^2$ determines the variations of the rescaled length $L_\text{RL}/L_0$ of the complexes, and an increase of this stiffness reduces the rotational phase space volume $\Omega_\text{RL}(l)$ of the complexes for a fixed local separation $l$ of the membranes. 
{
Changes in the anchoring strength $k_a$ of the receptors and ligands 
strongly affect the rotational free energy change $\Delta G_\text{rot}$ during binding. With decreasing  $k_a$, the effective width $\xi_\text{RL}$ of the function $K_\text{2D}(l)$ increases because the tilting of the complexes at small separations $l$ is facilitated (see Eq.\ (\ref{xiRL})). The decrease of the maximum value of the function $K_\text{2D}(l)$ with decreasing $k_a$ reflects that a more flexible anchoring of receptors and ligands for smaller values of $k_a$ results in a larger loss of rotational entropy upon binding and, thus, a larger rotational free energy change $\Delta G_\text{rot}$.}

\begin{figure}[t]
\resizebox{\columnwidth}{!}{\includegraphics{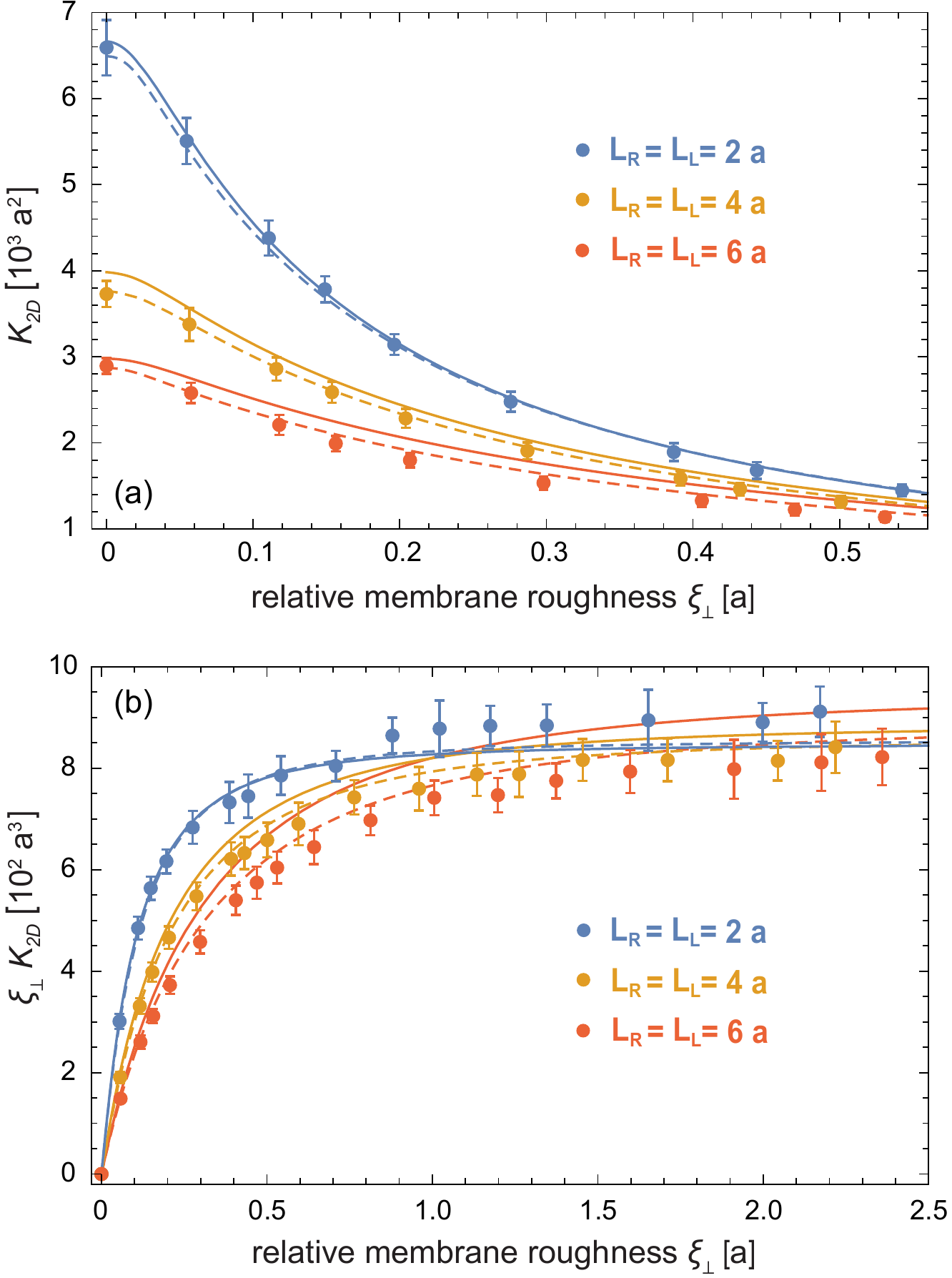}}
%\resizebox{0.5\columnwidth}{!}{\includegraphics{figure3}}
\caption{Binding constant $K_\text{2D}$ versus relative membrane roughness $\xi_\perp$ for the anchoring strength $k_a = 8 \, k_B T$ and different lengths $L_R = L_L$ of the receptors and ligands.  The full lines result from Eq.\ (\ref{K2D}) with the theoretical results for $K_\text{2D}(l)$ shown as full lines in Fig.\ \ref{figure_MC-planar}(a). The dashed lines are calculated based on the dashed interpolation lines of Fig.\ \ref{figure_MC-planar}(a). The left-most MC data points for zero roughness $\xi_\perp$ correspond to the maxima of the curves $K_\text{2D}(l)$ in Fig.\ \ref{figure_MC-planar}(a). In (b), the 11 points on the right results from MC simulations with flexible membranes of zero tension with $N_R=N_L=50$, 60, 75, 100, 125, 150, 200, 300, 500, 750, 1000 receptors and ligands (from right to left). The points 2 to 6 from left are from simulations with tension $\sigma = 500$, 100, 50, 25, 10 $k_BT/a^2$ and $N_R=N_L=100$. For $L_R = L_L=4\,a$ and $6\,a$,  point 7 from left results from simulations with tension $\sigma = 4\,  k_BT/a^2$. The data in (a) are from the same simulations as in the scaling plot (b), but for clarity shown only for a smaller roughness range. 
}
\label{figure_MC-fluc-ka8}
\end{figure}
\begin{figure}[t]
\resizebox{\columnwidth}{!}{\includegraphics{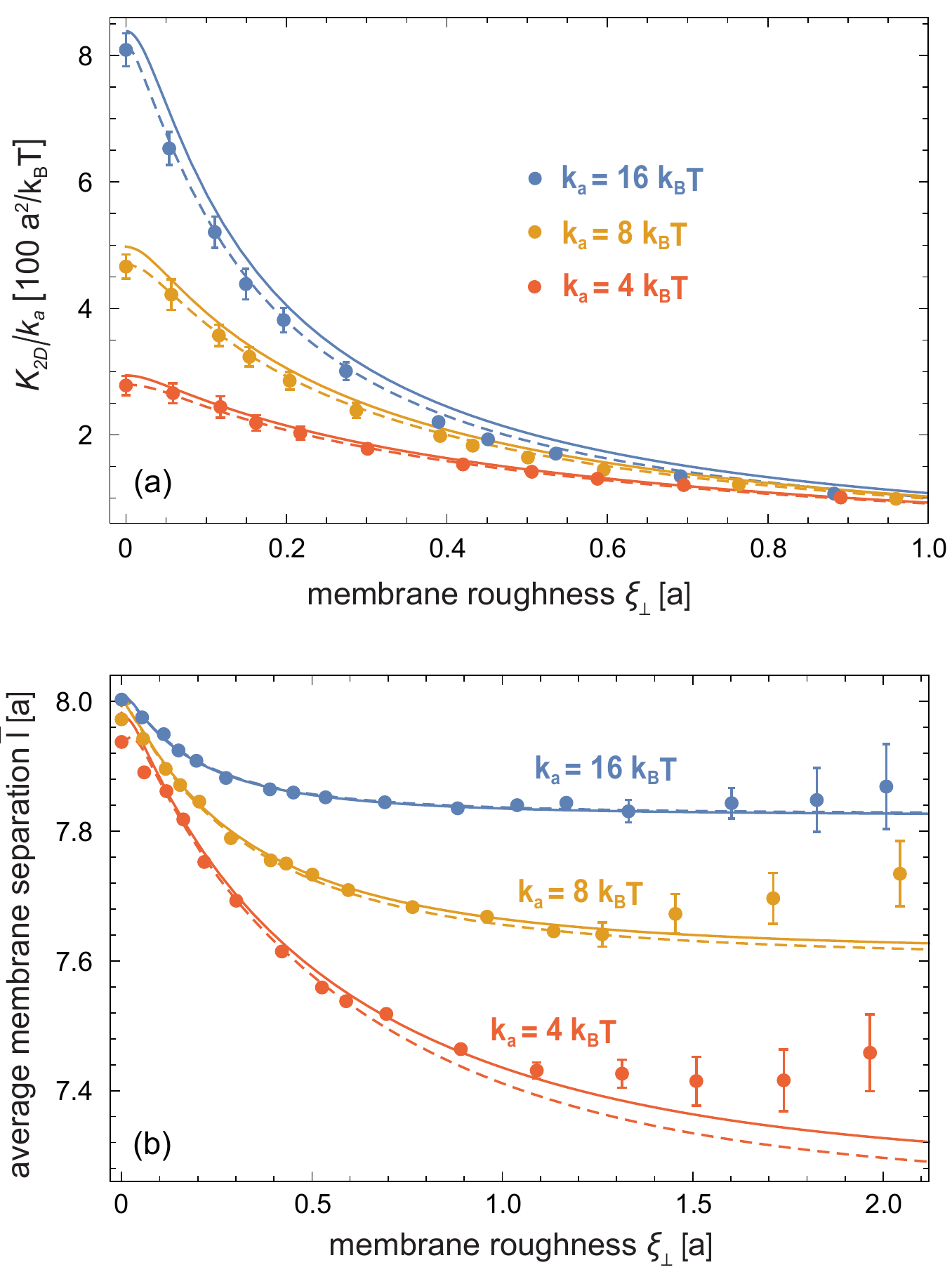}}
%\resizebox{0.6\columnwidth}{!}{\includegraphics{figure4}}
\caption{Rescaled binding constant $K_\text{2D}/k_a$ and average membrane separation $\bar{l}$ versus relative membrane roughness $\xi_\perp$ for the length $L_R = L_L= 4 \, a$ and different anchoring strengths $k_a$ of receptors and ligands. The full lines in this figure are obtained from Eq.\ (\ref{K2D}) and the theoretical results for $K_\text{2D}(l)$ shown as full lines in Fig.\ \ref{figure_MC-planar}(b). The dashed lines result from the dashed interpolation lines of Fig.\ \ref{figure_MC-planar}(b) (see text for details). The left-most MC data points correspond to the value and location of the maxima  in Fig.\ \ref{figure_MC-planar}(b). All other data points result from MC simulations with flexible membranes (see Fig.\ \ref{figure_MCsnapshots}(b)) for different membrane tensions $\sigma$ and numbers $N_R = N_L$ of receptors and ligands.}
\label{figure_MC-fluc-L4}
\end{figure}
\subsection{Binding constant of rigid receptors and ligands anchored to thermally rough membranes.}  

In our MC simulations with flexible membranes, the two membranes exhibit a relative roughness $\xi_\perp$ that results from thermally excited membrane shape fluctuations, and are `free to choose'  an optimal average separation $\bar{l}_0$ at which the overall free energy is minimal. In Figs.\ \ref{figure_MC-fluc-ka8} and \ref{figure_MC-fluc-L4}, MC data from these simulations are compared to our theory. The full lines in these figures are calculated from averaging our theoretical results for $K_\text{2D}(l)$ over the local membrane separation $l$ according to Eq.\ (\ref{K2D}), and do not involve any fit parameters. In this calculation, we approximate the distribution $P(l)$ of the local membrane separation $l$, which reflects the membrane shape fluctuations, by the Gaussian distribution  (\ref{pl}), and choose the average separation $\bar{l}$ of this distribution such that the binding constant $K_\text{2D}$ of Eq.\  (\ref{K2D}) is maximal, because maxima of $K_\text{2D}$ correspond to minima of the overall binding free energy of the adhering membranes. The width of the distribution $P(l)$ is the relative membrane roughness $\xi_\perp$. The dashed lines in the Figs.\ \ref{figure_MC-fluc-ka8} and \ref{figure_MC-fluc-L4} are calculated with the dashed interpolation functions for $K_\text{2D}(l)$ from Fig.\ \ref{figure_MC-planar}. 

The Figs.\ \ref{figure_MC-fluc-ka8}(a) and \ref{figure_MC-fluc-L4}(a) illustrate that the binding constant $K_\text{2D}$ decreases with increasing relative roughness $\xi_\perp$ of the membranes. The full theory lines in these figures do not involve any data fitting and agree overall well with the MC data. Slight deviations between the MC data and theory appear to result predominantly from a slight overestimation of the function $K_\text{2D}(l)$ in our theory (see Fig.\ \ref{figure_MC-planar}). The average over local separations of Eq.\ (\ref{K2D}) with the Gaussian approximation (\ref{pl}) does not seem to contribute significantly to these slight deviations, because the dashed lines in Figs.\ \ref{figure_MC-fluc-ka8}(a) and \ref{figure_MC-fluc-L4}(a) tend to agree with the MC data within statistical errors. These dashed lines are calculated based on the dashed interpolations of the MC data for $K_\text{2D}(l)$ in Fig.\ \ref{figure_MC-planar}.

For roughnesses $\xi_\perp$ that are much larger than the effective width $\xi_\text{RL}$ of the functions $K_\text{2D}(l)$ shown in Fig.\ \ref{figure_MC-planar}, the binding constant $K_\text{2D}$ is inversely proportional to $\xi_\perp$ at the optimal average separation $\bar{l} = \bar{l}_0$ for binding (see Eq.\ (\ref{K2Dlim})). In the scaling plot of Fig.\ \ref{figure_MC-fluc-ka8}(b), $\xi_\perp K_\text{2D}$ therefore tends to constant, limiting values for large roughnesses $\xi_\perp\gg \xi_\text{RL}$. Based on Eq.\ (\ref{xiRL}), the effective width of the function $K_\text{2D}(l)$ can be estimated as $\xi_\text{RL}\simeq  0.26\,a$, $0.51\,a$, and $0.75a\,$ for the receptor and ligand lengths $L_R=L_L= 2\,a$, $4\,a$, and $6\,a$ of Fig.\ \ref{figure_MC-fluc-ka8}(b). Because of the smaller value of $\xi_\text{RL}$, the blue curve in Fig.\ \ref{figure_MC-fluc-ka8}(b) for the receptor and ligand length $L_R=L_L= 2\,a$ approaches its limiting value faster than the other two curves.

Fig.\  \ref{figure_MC-fluc-L4}(b) illustrates that the preferred average separation of the two adhering membranes decreases with the relative roughness $\xi_\perp$ of the membranes. The lines in this figure result from maximizing $K_\text{2D}$ in Eq.\ (\ref{K2D}) with respect to the average separation of the Gaussian distribution $(\ref{pl})$ for the functions $K_\text{2D}(l)$ shown in Fig.\ \ref{figure_MC-planar}(b). The full lines are based on our theoretical calculations of $K_\text{2D}(l)$ and do not involve any data fitting. The dashed lines are calculated based on the dashed interpolations of the MC data for $K_\text{2D}(l)$ in Fig.\ \ref{figure_MC-planar}(b). For small and intermediate roughnesses, the lines in Fig.\ \ref{figure_MC-fluc-L4}(b) agree well with the data points from our MC simulations in which the membranes can `freely choose' a preferred average separation $\bar{l}_0$ . For large roughnesses, the MC data deviate from the theory lines because of the fluctuation-induced repulsion of the impenetrable membranes, which is not taken into account in our theory. In the roughness range in which the fluctuation-induced repulsion of the membranes is negligible, the preferred average separation $\bar{l}_0$ decreases because of the asymmetry of the function $K_\text{2D}(l)$. At zero roughness, the preferred average separation $\bar{l}_0$ is identical to the local separation $l_0$ at which $K_\text{2D}(l)$ is maximal. For larger roughnesses, the average of $K_\text{2D}(l)$ over the local separations $l$ in Eq.\ (\ref{K2D}) is maximal at average separations $\bar{l}_0$ smaller than $l_0$ because $K_\text{2D}(l)$ is asymmetric, with a pronounced `left arm' that reflects tilting of the receptor-ligand complexes. The preferred average separation $\bar{l}_0$ decreases for decreasing anchoring strength $k_a$ because of smaller tilt energies. For roughnesses $\xi_\perp$ that are large compared to the width $\xi_\text{RL}$ of the functions $K_\text{2D}(l)$, the preferred average separation $\bar{l}_0$ in our theory can be estimated from Eq.\ (\ref{barl0}), which leads to $\bar{l}_0 = 7.06\,a$, $7.56\,a$,  and $7.81\,a$ for the anchoring strengths $k_a = 4\, k_B T$, $8\, k_B T$, and $16\, k_B T$ of Fig.\ \ref{figure_MC-fluc-L4}(b) and the preferred length $L_0 \simeq 8.07\, a$ of the receptor-ligand complex with the molecular lengths $L_R=L_L= 4a$. 

\begin{figure*}[t]
\resizebox{1.6\columnwidth}{!}{\includegraphics{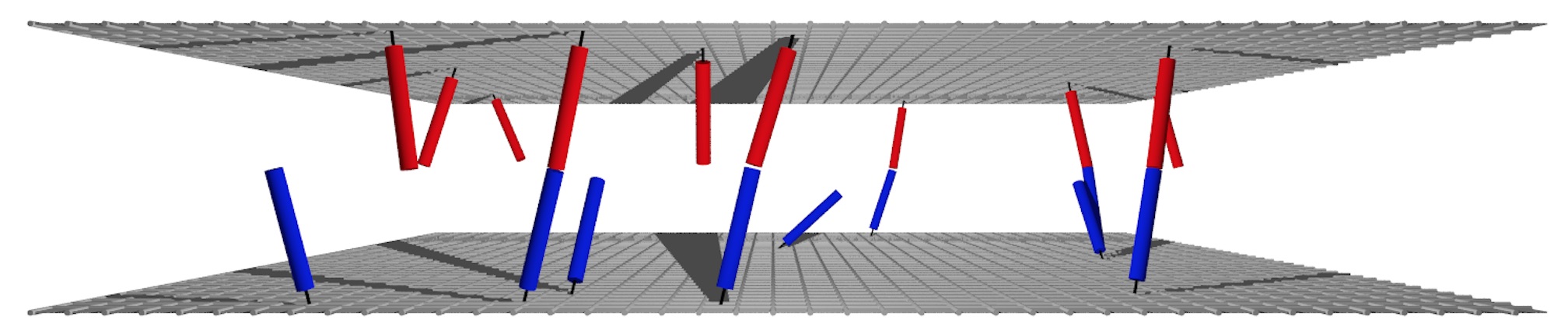}}
%\resizebox{\columnwidth}{!}{\includegraphics{figure5}}
\caption{Snapshot from a MC simulation with semi-flexible receptors and ligands anchored to parallel and planar membranes. In this simulation, the membrane separation is $l = 7.8\,a$, and the receptors and ligands have an anchoring strength $k_a=8\, k_BT$ and stiffness  $k_f=64\, k_BT$. Each semi-flexible receptor or ligand is composed of two rod-like segments of length $2\,a$, which are connected by a joint with stiffness $k_f$. The snapshot displays segments with area $40 \times 40\,a^2$ of the simulated membranes, which have the overall area $160 \times 160\,a^2$. 
 }
\label{figure_MCsnapshot-flexible}
\end{figure*}
\begin{figure}[t]
\resizebox{\columnwidth}{!}{\includegraphics{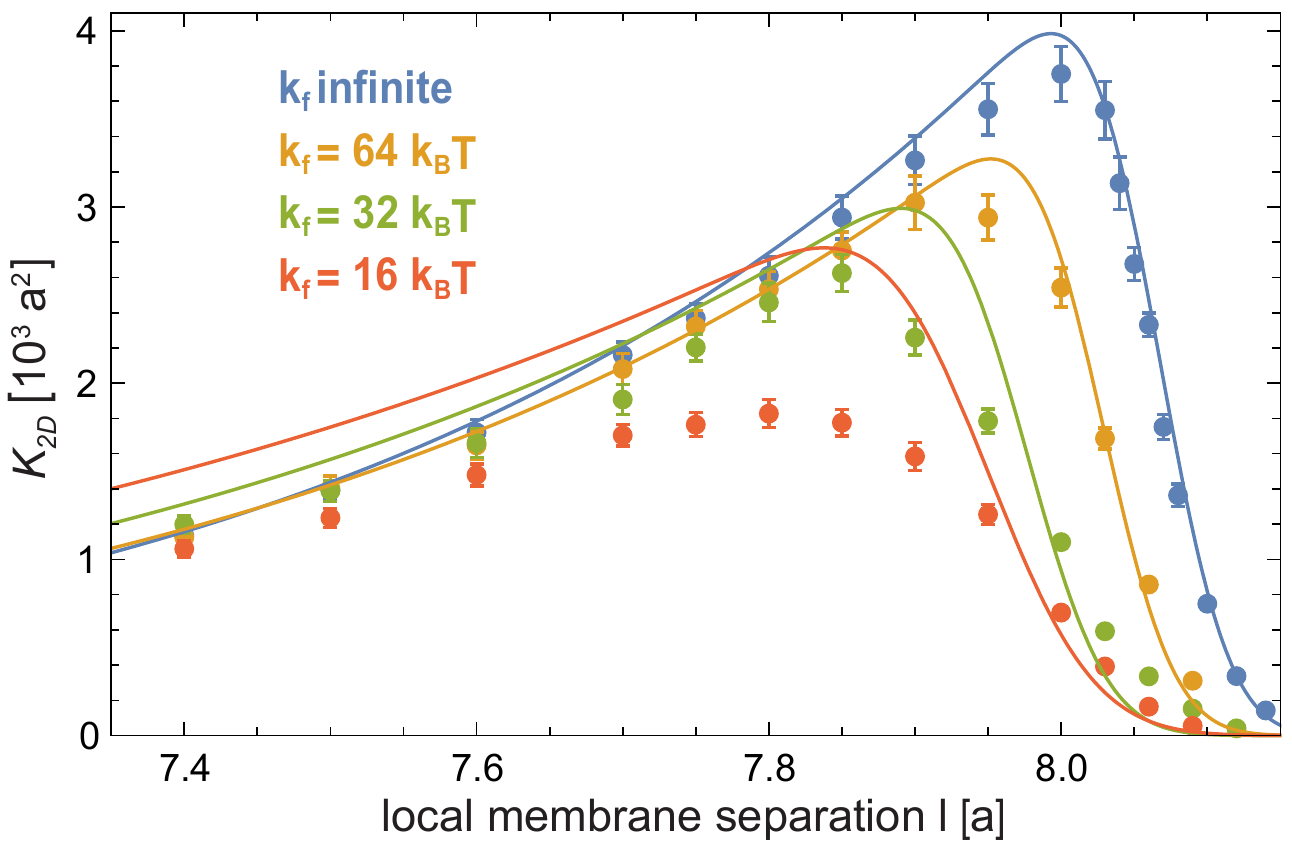}}
%\resizebox{0.6\columnwidth}{!}{\includegraphics{figure6}}
\caption{Binding constant $K_\text{2D}$ of semi-flexible receptor and ligand molecules as a function of the local separation $l$ for different stiffnesses $k_f$ of the molecules. The MC data points are from simulations with parallel and planar membranes. The anchoring strength of the receptors and ligands in these simulations is $k_a = 8\, k_B T$, and the length of the two rod-like segments of the semi-flexible receptors and ligands is $2\,a$. The lines in Fig.\ \ref{figure_MC-flex} represent theoretical results based on Eq.\ (\ref{K2Dl}) with the effective anchoring strengths $k_a^\text{eff}= 5.25\, k_BT$, $6.34\, k_B T$, and $7.08\,k_B T$ and spring constants $k_\text{RL} = 295 \; k_BT/a^2$, $528 \; k_BT/a^2$, and $664\, k_BT/a^2$ for $k_f = 16 \,k_B T$, $32\,k_B T$, and $64\,k_B T$, respectively. The values $L_0 = 7.95\,a$, $7.97\,a$, and $8.03\,a$ for the preferred length of the RL complexes with stiffness $k_f = 16 \,k_B T$, $32\,k_B T$, and $64\,k_B T$ here are obtained from a fit to the MC data. The blue data points and lines for infinite stiffness $k_f$ correspond to the yellow data points and lines for rod-like receptors and ligands in Fig.\ \ref{figure_MC-planar} with anchoring strength $k_a = 8\, k_B T$ and length $L_R = L_L = 4\,a$.
}
\label{figure_MC-flex}
\end{figure}
\subsection{MC data and theory for the binding constants of semi-flexible receptors and ligands}

In this section, we extend our theory to semi-flexible receptors and ligands and compare this extended theory to MC data. Each semi-flexible receptor and ligand in our MC simulations consist of two rod-like segments, an anchoring segment and an interacting segment, that are connected by a flexible joint with bending energy 
\begin{equation}
V_\text{ben}=\frac {1}{2} k_f \theta_f^2
\label{Vben}
\end{equation}
and stiffness $k_f$ (see also Fig.\ \ref{figure_MCsnapshot-flexible}). The overall configurational energy (\ref{overall_energy}) then contains the total bending energy $H_\text{ben}$ of all receptors and ligands as an additional term. As additional type of MC move, our simulations with semi-flexible receptor and ligand molecules involve continuous rotational moves around the flexible joints connecting the two rod-like segments of the molecules. The anchoring segment of a semi-flexible receptor or ligand is attached to the membrane via the same anchoring potential (\ref{Vanchor}) as the rod-like receptors and ligands. The interacting segments of a semi-flexible receptor and ligand interact via the same binding potential (\ref{binding_potential}). Since the binding constant $K_\text{3D}$ of soluble receptors and ligands only depends on the binding potential, our semi-flexible receptors and ligands have the same value of $K_\text{3D}$ as our rod-like receptors and ligands, irrespective of their stiffness $k_f$. 

In contrast, {  the maximum value of the binding constant $K_\text{2D}$} of membrane-anchored semi-flexible receptors and ligands decreases with decreasing stiffness $k_f$ (see Fig.\ \ref{figure_MC-flex}). The MC data for $K_\text{2D}(l)$ in this figure result from simulations with parallel and planar membranes (see Fig.\ \ref{figure_MCsnapshot-flexible}). In these simulations, both rod-like segments of a receptor or ligand have the length $2\,a$, and the anchoring segment is anchored to the membrane with strength $k_a = 8\, k_B T$. We consider semi-flexible receptors and ligands with the three different stiffness values $k_f = 16 \, k_B T$, $32 \, k_B T$, and $64 \, k_B T$. An infinite stiffness $k_f$ corresponds to rod-like receptors and ligands with length $4\,a$. The blue data in Fig.\ \ref{figure_MCsnapshot-flexible} for infinite $k_f$ therefore correspond to the yellow data of Fig. \ref{figure_MC-planar} for $k_a = 8\, k_B T$ and $L_R = L_L = 4\,a$.

We find that the function $K_\text{2D}(l)$ for the semi-flexible receptor and ligand molecules can be described for large stiffness $k_f\gg k_a$ by a reduced effective anchoring strength $k_a^\text{eff}$ in our theory for rod-like molecules. This effective anchoring strength can be calculated from the standard deviation of the angle $\theta_i$ of the interacting segment of the semi-flexible molecules with respect to the membrane normal. For the anchoring strength $k_a = 8\,k_B T$ as in Fig.\ \ref{figure_MC-flex}, the standard deviation of the angle $\theta_i$ is 0.597 for $k_f=16 k_B T$,  0.547 for $k_f=32 k_B T$, 0.519 for  $k_f=64\,k_B T$, and 0.489 for infinite $k_f$, which corresponds to rod-like receptors and ligands with $\theta_i = \theta_a$.  We obtain the same standard deviations for the angle $\theta_a$ of rod-like molecules with the effective anchoring strengths $k_a^\text{eff}= 5.25\, k_BT$, $6.34\, k_B T$, and $7.08\,k_B T$ for $k_f = 16 \,k_B T$, $32\,k_B T$, and $64\,k_B T$, respectively. The lines in Fig.\ \ref{figure_MC-flex} represent our theoretical results based on Eq.\ (\ref{K2Dl}) for rod-like molecules with these effective anchoring strengths and with the values $k_\text{RL} = 295 \; k_BT/a^2$, $528 \; k_BT/a^2$, and $664\, k_BT/a^2$ for $k_f = 16 \,k_B T$, $32\,k_B T$, and $64\,k_B T$, respectively, which are obtained from the standard deviations of the end-to-end distance determined in MC simulations of soluble RL complexes. The preferred length $L_0$ of the semi-flexible RL complexes are obtained from a fit to the MC data in Fig.\ \ref{figure_MC-flex}. The theoretical results for $K_\text{2D}(l)$ are in good agreement with the MC data for the stiffness $k_f = 64\,k_BT$, which is much larger than the anchoring strength $k_a= 8\, k_B T$. For the smaller stiffnesses $k_f = 32\,k_BT$ and $16\,k_BT$, the theoretical results deviate more strongly from the MC data, which indicates that our extended theory based on effective anchoring strengths is valid for $k_f \gg k_a$. 

%%%
\section{Discussion and conclusions}
%%%

We have presented here a general theory for the binding equilibrium constant $K_\text{2D}$ of rather stiff membrane-anchored receptors and ligands. This theory generalizes our previous theoretical results \cite{Hu13}  by describing how $K_\text{2D}$ depends both on the average separation $\bar{l}$ and thermal nanoscale roughness $\xi_\perp$ of the apposing membranes, and on the anchoring, length and flexibility of the receptors and ligands.  A central element of this theory is the calculation of the rotational phase space volume  of the bound receptor-ligand complex, which is based on an effective configurational energy of the complex (see Eqs.\ (\ref{Hef}) to (\ref{OmegaRL})). In our previous theory for the preferred average membrane separation $\bar{l}_0$ for binding, the rotational phase space volume of the bound complex was determined from the distribution of anchoring angles of the complex observed in simulations \cite{Hu13}.  In the theory presented here, the dependence of $K_\text{2D}$ on the average membrane separation $\bar{l}$ and relative roughness $\xi_\perp$ results from averaging $K_\text{2D}(l)$ over the distribution $P(l)$ of local membrane separations $l$ with mean $\bar{l}$ and standard deviation $\xi_\perp$. For relative roughnesses $\xi_\perp$ that are much larger than the the width of the function$K_\text{2D}(l)$, the binding constant $K_\text{2D}$ is inversely proportional to $\xi_\perp$ at average membrane separations $\bar{l}$ equal to the preferred average separation $\bar{l}_0$ according to Eq.\ (\ref{K2Dlim}). In our previous theory, this inverse proportionality resulted from the entropy loss of the membranes upon receptor-ligand binding. Our theories relate the binding constant $K_\text{2D}$ of the membrane-anchored receptor and ligand proteins to the binding constant $K_\text{3D}$ of soluble variants of the proteins without membrane anchors by determining the translational and rotational free energy changes of anchored and soluble proteins upon binding. In a complementary approach of Wu et al.\cite{Wu10,Wu13}, the binding constant  $K_\text{2D}$ of receptors and ligands anchored to essentially planar membranes is determined based on ranges of motion of bound and unbound receptors and ligands in the direction perpendicular to the membranes. 

In this article, we have corroborated our theory by a comparison to detailed data from MC simulations. Our general results for the ratio  $K_\text{2D}/K_\text{3D}$ of the binding constants of membrane-anchored and soluble receptors and ligands agree with the MC results without any data fitting. Our MC simulations are based on a novel elastic-membrane model in which the receptors and ligands are described as anchored molecules that diffuse continuously along the membranes and rotate at their anchoring points. In our accompanying article\cite{Hu}, we compare our general theoretical results for $K_\text{2D}$ to detailed data from molecular dynamics simulations of biomembrane adhesion with both transmembrane and lipid-anchored receptors and ligands, and extend our theory to the binding rate constants $k_\text{on}$ and  $k_\text{off}$.   Our theoretical results are rather general and hold for membrane-anchored molecules whose anchoring is `soft' compared to their binding and bending, which is realistic for a large variety of biologically important membrane receptors and ligands such as the T-cell receptor and its MHC-peptide ligand or the cell adhesion proteins CD2, CD48, and CD58. 

The dependence of the binding constant $K_\text{2D}$ on the average separation $\bar{l}$ and relative roughness $\xi_\perp$ of the membranes helps to understand why { mechanical} methods that probe the binding kinetics of membrane-anchored proteins during initial membrane contacts can lead to values for the binding equilibrium constant $K_\text{2D}$ that are orders of magnitude smaller than the values obtained from { fluorescence measurements} in equilibrated adhesion zones \cite{Dustin01}. In equilibrated adhesion zones that are dominated by a single species of receptors and ligands, the average membrane separation $\bar{l}$ is close to the preferred average separation $\bar{l}_0$ for binding, and the relative membrane roughness $\xi_\perp$ is reduced by receptor-ligand bonds \cite{Krobath09,Hu13}. During initial membrane contacts, in contrast, both the membrane separation $\bar{l}$ and roughness $\xi_\perp$  are larger, which can lead to significantly smaller values for $K_\text{2D}$ according to our theory. 

{
In our MC simulations, we have focused on membranes that adhere {\em via} a single species of receptors and ligands. The average membrane separation $\bar{l}$ then is identical to the preferred average separation $\bar{l}_0$ of these receptors and ligands for binding. However, our elastic-membrane model can be generalized to situations in which membrane adhesion is mediated by different species of receptors or ligands, e.g.\ by long and short pairs of receptors or ligands as in T-cell adhesion zones \cite{Monks98,Grakoui99,Mossman05}, or to situations in which the binding of receptors and ligands is opposed by repulsive membrane-anchored molecules, e.g.\ by molecules of the cellular glycocalyx \cite{Paszek14}. These situations have been previously investigated with elastic-membrane models in which the molecular interactions of receptors and ligands or repulsive molecules are described implicitly by interaction potentials that depend on the local membrane separation \cite{Weikl01,Qi01,Weikl02a,Chen03,Raychaudhuri03,Weikl04,Asfaw06,Coombs04,Wu06}. At sufficiently large concentrations, long and short receptor and ligand molecules segregate into domains in which the adhesion is dominated either by the short or by the long molecules \cite{Weikl09,Rozycki10}. The domain formation is caused by a membrane-mediated repulsion between long and short receptor-ligand complexes, which arises from membrane bending to compensate the length mismatch.
In each domain, the average separation of the membranes is close to the preferred average separation of the dominating receptors and ligands. Within such a domain, the distribution $P(l)$ of the local membrane separation $l$ has a single peak centered around the preferred average separation of the dominating receptors and ligands. Averaged over whole adhesion zones with multiple domains, the distribution $P(l)$ has two peaks that are centered around the preferred average separations of the long and short pairs of receptors and ligands. 
%At small concentrations, long and short receptors and ligands can mix in adhesion zones, which leads to average membrane separations  
Similarly, short receptor and ligand molecules and longer repulsive molecules segregate at sufficiently large molecular concentrations \cite{Weikl01,Weikl02a,Paszek09}. 
}

 { Several groups have investigated experimentally how varying the length of membrane-anchored receptors or ligands affects cell adhesion. Chan and Springer  \cite{Chan92} found an increased cell-cell adhesion efficiency in hydrodynamic flow for elongated variants of CD58, compared to wild-type CD58. Patel et al.\  \cite{Patel95} observed that cells with long variants of P-selectin bind more efficiently under shear flow to cells with the binding partner PSGL-1, compared to shorter variants of P-selectin. %In the absence of flow, in contrast, Patel el al.\ observed equivalent adhesion efficiencies of short and long variants of P-selectin. 
From adhesion frequencies in a micropipette setup, Huang et al.\ \cite{Huang04} obtained higher on-rates for long P-selectin constructs attached to red-blood-cell surfaces, compared to short P-selectin constructs, and identical off-rates for both constructs. These results indicate that initial cell-cell adhesion events probed in hydrodynamic flow or with micropipette setups can be more efficient for elongated receptors or ligands, presumably  due to reduced cytoskeletal repulsion \cite{Chan92,Patel95}.
In a different approach, Milstein et al.\ \cite{Milstein08} investigated the CD2-mediated adhesion efficiency of T cells to supported membranes that contain either wild type CD48 or elongated variants of CD48. For elongated variants of CD48, Milstein et al.\ observed less efficient cell adhesion after one hour compared to wild type CD48 at identical concentrations. This observation is in qualitative agreement with our findings that the binding constant $K_\text{2D}$ decreases with increasing length of receptors and ligands (see Fig.\ 3(a)), and increasing flexibility (see Fig.\ 6). Besides increasing the length, the addition of protein domains may lead to a larger flexibility of the elongated variants of CD48 compared to the wildtype. 
}

{
We have focused here on receptors and ligands with preferred collinear binding and preferred perpendicular membrane anchoring, i.e.\ with a preferred anchoring angle of zero relative to the membrane normal. A preferred non-zero anchoring angle $\theta_0$ can be simply taken into account by changing the anchoring energy (\ref{Vanchor}) to $V_\text{anchor} = \frac{1}{2} k_a (\theta_a - \theta_0)^2$. For a preferred collinear binding of rod-like receptors and ligands, the preferred binding angle $\theta_b$ is 0. For receptors and ligands anchored to parallel and planar membranes as in sections III.B and III.C, the anchoring angles of a receptor and ligand in a bound complex then are identical, and identical to the tilt angle $\theta_c$ of the receptor-ligand complex. The tilt angle $\theta_c$ here is defined as the angle between the membrane normal and the line connecting the two anchor points of the receptor-ligand complex. For a preferred non-zero binding angle $\theta_b$, the receptor-ligand complex is kinked. The anchoring angles $\theta_{a,1}$ and $\theta_{a,2}$ 
of a receptor and ligand in a bound complex then depend not only on the tilt angle $\theta_c$ of the complex, but also on the torsional angle $\phi_c$ of the complex around the tilt axis, the lengths $L_R$ and $L_L$ of the receptor and ligand, and the preferred binding angle. The rotational phase space volume $\Omega_\text{RL}$ of such a kinked RL complex can be calculated by integrating $\exp(-H_\text{RL}/k_B T)$ over the tilt angle $\theta_c$ and torsional angle $\phi_c$ of the complex, where $H_\text{RL}$ is the generalized effective configurational energy of the complex with anchoring angles  $\theta_{a,1}$ and $\theta_{a,2}$.

The rod-like receptors and ligands and rod-like segments of semi-flexible receptors and ligands considered here can freely rotate around their axes in the bound and unbound state. For proteins, in contrast, such rotations will be restricted in the bound complex, which leads to an additional loss of rotational entropy upon binding. However, this additional loss of rotational entropy is identical both for the membrane-anchored complex in 2D and the soluble complex in 3D and, thus, does not affect the ratio $K_\text{2D}/K_\text{3D}$ of the binding constants, provided the binding interface of the receptor-ligand complex is not affected by membrane anchoring, as assumed in section III.C.}

\appendix

\begin{figure}[t]
\resizebox{0.6\columnwidth}{!}{\includegraphics{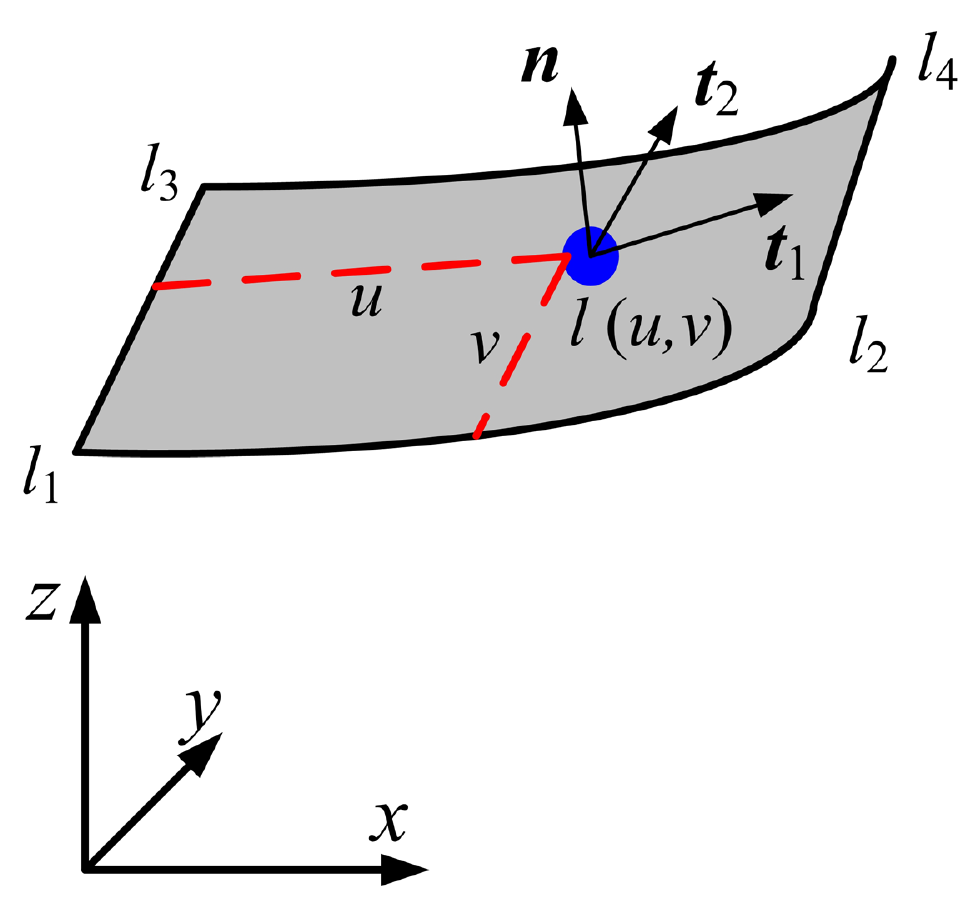}}
%\resizebox{0.4\columnwidth}{!}{\includegraphics{figure7}}
\caption{Illustration of a quadratic membrane patch with a receptor or ligand located at $(u,v)$. The local deviation $l(u,v)$ and the tangent vectors $\boldsymbol{t}_1$ and $\boldsymbol{t}_2$ and normal vector $\boldsymbol{n}$ follow from linear interpolation (see text).
}
\label{LinearInter}
\end{figure}
\section{Positions and anchoring angles of receptors and ligands in our elastic-membrane model.}

In our elastic-membrane model of biomembrane adhesion, the conformations of the two apposing membranes are described by local deviations $l_i$ at lattice sites $i$ of a reference plane. The receptors and ligands of this model move continuously along the membranes and, thus, `in between' the discretization sites of the membrane. The anchor position and anchoring angle of a receptor or ligand can be obtained by linear interpolation from the local membrane deviations $l_1$, $l_2$, $l_3$, and $l_4$ at the four lattice sites 1, 2, 3, and 4 around the receptor or ligand (see Fig.\ \ref{LinearInter}). The anchor position of the receptor or ligand within a quadratic patch of the reference plane with corners 1, 2, 3, and 4 can be described by the parameters $u$ and $v$ with $0 \leq u,v \leq 1$. The local membrane deviation $l(u,v)$ of the anchor out of the reference plane then follows from linear interpolation \cite{Reister07,Naji09}:
\begin{equation}
l(u,v)=(1-u)(1-v)l_1+u(1-v)l_2+(1-u)vl_3+uvl_4
\end{equation}

\begin{figure}[t]
\resizebox{0.6\columnwidth}{!}{\includegraphics{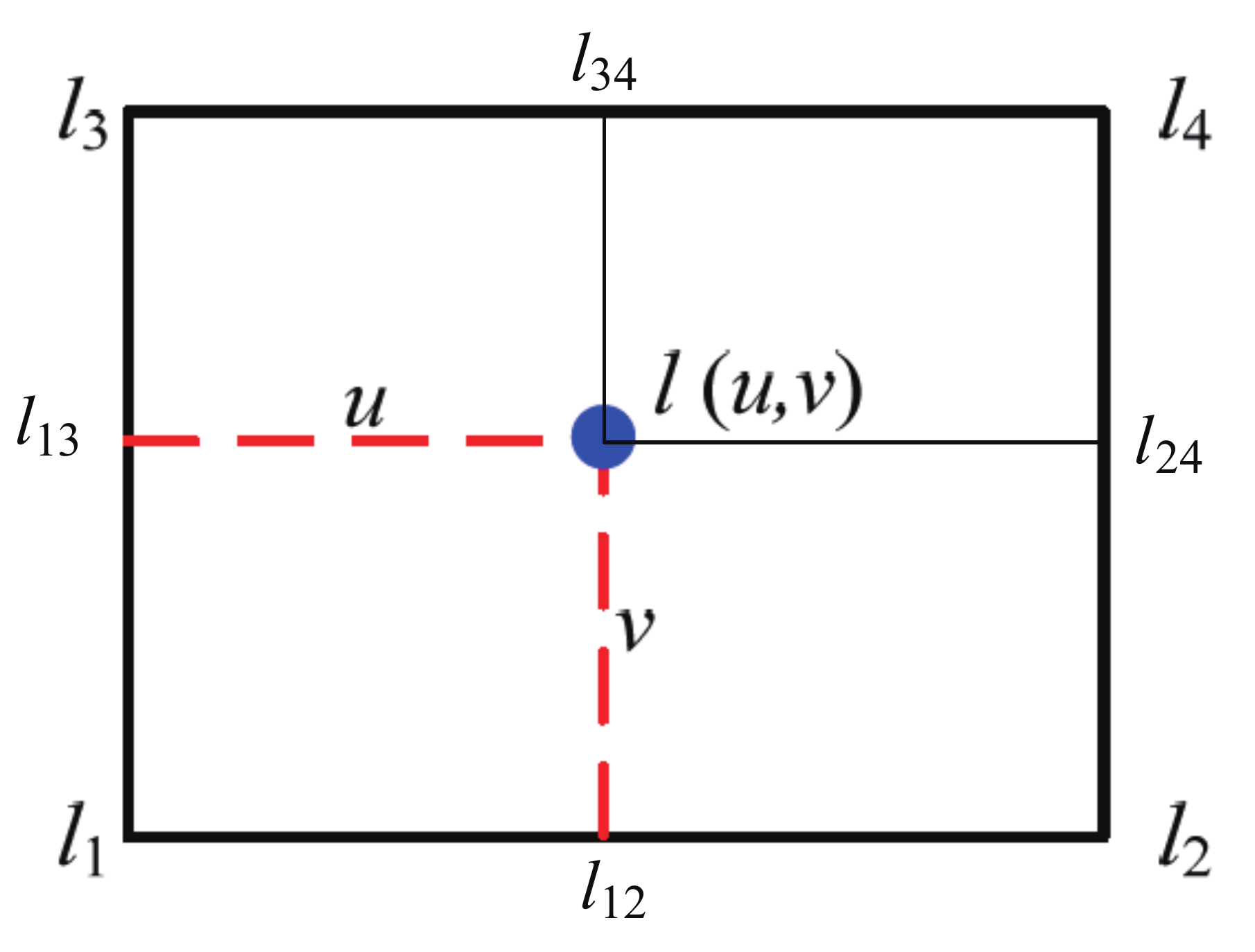}}
%\resizebox{0.4\columnwidth}{!}{\includegraphics{figure8}}
\caption{Projection of the quadratic membrane patch shown in Fig.\ \ref{LinearInter}.}
\label{Planform}
\end{figure}

To calculate the anchoring angle of a receptor or ligand molecule, we first need to determine the membrane normal $\boldsymbol{n}$ at the site $(u,v)$ of the anchor. The membrane normal can be calculated from the two tangent vectors $\boldsymbol{t}_1$ and $\boldsymbol{t}_2$ of the membrane at site $(u,v)$ (see Fig.\ \ref{LinearInter}). The tangent vector $\boldsymbol{t}_1$ is
\begin{equation}
\boldsymbol{t}_1=\cos \theta_x \boldsymbol{e}_x + \sin \theta_x \boldsymbol{e}_z
\end{equation}
where $\boldsymbol{e}_x$ and $\boldsymbol{e}_z$ are the unit vectors along the $x$ and $z$ axis. The angle $\theta_x$ between the vector $\boldsymbol{t}_1$ and the $x$ axis can be obtained from
\begin{equation}
\tan \theta_x=l_{24}-l_{13}=(1-v)(l_2-l_1)+v(l_4-l_3)
\end{equation}
with $l_{24}$ and $l_{13}$ illustrated in Fig.\ \ref{Planform}. For simplicity, all lengths here are normalized by the lattice spacing $a$. Similarly, the tangent vector $\boldsymbol{t}_2$ is
\begin{equation}
\boldsymbol{t}_2=\cos \theta_y \boldsymbol{e}_y+\sin \theta_y \boldsymbol{e}_z
\end{equation}
where the angle $\theta_y$ between the vector $\boldsymbol{t}_2$ and the $y$ axis can be obtained from
\begin{equation}
\tan \theta_y=l_{34}-l_{12}=(1-u)(l_3-l_1)+u(l_4-l_2)
\end{equation}
From $\boldsymbol{t}_1$ and $\boldsymbol{t}_2$, the membrane normal vector $\boldsymbol{n}$ can be calculated as
\begin{equation}
\boldsymbol{n}=\boldsymbol{t}_1 \times \boldsymbol{t}_2
\end{equation}
The anchoring angle $\theta_a$ between the rod-like receptor or ligand and the membrane normal then follows as
\begin{equation}
\theta_a=\arccos(\boldsymbol{n} \cdot \boldsymbol{r})
\end{equation}
where $\boldsymbol{r}$ is a unit vector pointing in the direction of the receptor or ligand.

\section{Effective configurational energy of receptor-ligand complexes.}

In this section, we derive the effective configurational energy (\ref{Hef}) of a receptor-ligand complex and Eqs.\ (\ref{L0}) and (\ref{kRL}) for the preferred length $L_0$ and the effective spring constant $k_\text{RL}$ of the complex. The length $L_\text{RL}$ of a receptor-ligand complex is the distance between the two anchor points of the receptor and ligand. For rod-like receptors and ligands, variations in this length mainly result from variations in the binding angle $\theta_b$ and in the binding-site distance $z$ in the direction of the complex. For small binding angles $\theta_b$, variations of the binding-site distance in the two directions $x$ and $y$ perpendicular to the complex can be neglected. The length of the complex is then
\begin{equation}
L_\text{RL}(\theta_b,z) \simeq z+\sqrt{L_\text{R}^2 + L_\text{L}^2 + 2 L_\text{R} L_\text{L} \cos\theta_b} 
\label{Ldef}
\end{equation}
where $L_\text{R}$ and $L_\text{L}$ are the lengths of the receptor and ligand. In harmonic approximation, the variations in the binding angle $\theta_b$ and binding-site distance $z$ in the direction parallel to the complex can be described by the configurational energy
\begin{equation}
H(\theta_b,z) = \frac{1}{2} k_b \theta_b^2 + \frac{1}{2} k_z (z -z_0)^2
\end{equation}
where $k_b$ and $k_z$ are spring constants that are related to the standard deviations $\sigma_b$ and $\xi_z$ of the distributions for the binding angle $\theta_b$ and binding-site distance $z$ via $k_b = k_B T/\sigma_b^2$ and $k_z = k_B T/\xi_z^2$.
We assume now that $k_b$ is much larger than the thermal energy $k_B T$, which implies small binding angles $\theta_b$. From expanding Eq.\ (\ref{Ldef}) up to second order in $\theta_b$, we obtain the average length
\begin{equation}
L_0 = \langle L_{RL} \rangle \simeq z_0 + L_\text{R} + L_\text{L}  - \frac{k_B T}{k_b} \frac{L_\text{R} L_\text{L}}{L_\text{R} + L_\text{L}} 
\end{equation}
and the variance of the length
\begin{equation}
\xi_0^2 = \langle L_\text{RL}^2 \rangle - \langle L_\text{RL} \rangle ^2 \simeq 
\frac{k_B T}{k_z} +\left(\frac{k_B T}{k_b}\right)^2 \frac{L_\text{R}^2 L_\text{L}^2}{(L_\text{R} + L_\text{L})^2}
\end{equation}
to leading order in $k_B T/k_b$. The thermodynamic averages here are calculated as 
\begin{equation}
\langle \ldots\rangle =  \frac{\int_{-\infty}^{\infty}\int_0^{\pi/2} \ldots e^{-H(\theta_b,z)/k_B T}\sin \theta_b {\rm d}\theta_b {\rm d}z}{\int_{-\infty}^{\infty}\int_0^{\pi/2}  e^{-H(\theta_b,z)/k_B T}\sin \theta_b {\rm d}\theta_b {\rm d}z}
\end{equation}
The variations in the end-to-end distance $L_\text{RL}$ of the receptor-ligand complex then can be described by the second term of effective configurational energy (\ref{Hef}) with the effective spring constant $k_\text{RL} = k_B T/\xi_0^2$ (see Eq.\ (\ref{kRL})).

\section{Integrals and moments of the function $K_\text{2D}(l)$.}

The shape of the function $K_\text{2D}(l)$ introduced in Eq.\ (\ref{K2Dl}) is determined by $\Omega_\text{RL}(l)$, i.e.\ by the rotational phase space volume of the RL complex as a function of the local separation $l$. The mean value and standard deviation of $K_\text{2D}(l)$ therefore is identical to the mean value and standard deviation of $\Omega_\text{RL}(l)$. We first consider here the moments of $\Omega_\text{RL}(l)$. 
The zeroth moment is the integral 
\begin{align}
m_0 & =\int_0^\infty \Omega_\text{RL}(l) {\rm d}l \nonumber \\
&= 2 \pi\int_{0}^{\infty}  \left[ \int_0^{\pi/2}  e^{-H_\text{RL}(\theta_a,L_\text{RL}(\theta_a))/k_B T} \sin\theta_a{\rm d}\theta_a\right] {\rm d}l 
\label{moA}\\
&\simeq 
2 \pi \int_0^{\infty}  \left[ \int_{-\infty}^{\infty} e^{-H_\text{RL}(\theta_a,L_\text{RL}(\theta_a))/k_B T}   {\rm d}l \right]\sin\theta_a{\rm d}\theta_a 
\label{moB}\\
& \simeq \frac{\sqrt{2} \pi^{3/2} k_B T}{\sqrt{k_a k_\text{RL}}}  F_D\left(\sqrt{\frac{k_B T}{k_a}}\right)
\label{moC}
\end{align}
where $F_D$ is the Dawson function. The approximate result (\ref{moC}) holds for anchoring strengths $k_a\gg k_B T$ for which the integrand is practically 0 at the upper limit $\pi/2$ of the integration over $\theta_a$ in Eq.\ (\ref{moA}). This approximate result then is obtained by interchanging the order of the integrations over $\theta_a$ and $l$, and by extending integration limits to infinity. We assume that the binding interaction is rather `hard' compared to the anchoring, which implies $k_\text{RL} L_0^2\gg k_a$. In the same way, the first and second moment of $\Omega_\text{RL}(l)$ are obtained as

\begin{align}
m_1 &= \int_0^\infty l \;\Omega_\text{RL}(l) {\rm d}l \\
&\simeq \frac{\pi^{3/2}L_0 k_BT}{\sqrt{2 k_a k_\text{RL}}} 
\Bigg[F_D\left(\frac{1}{2}\sqrt{\frac{k_B T}{k_a}}\right) \nonumber \\
&\hspace{3cm}+ F_D\left(\frac{1}{2}\left(\frac{k_B T}{k_a}\right)^{3/2}\right) \Bigg]
\end{align}
and
\begin{align}
m_2 &= \int_0^\infty l^2 \;\Omega_\text{RL}(l) {\rm d}l \\
&\simeq \frac{\pi^{3/2}k_B T\left(k_\text{RL}L_0^2 + k_B T\right)}{\sqrt{8 k_a}k_\text{RL}^{3/2}}
\Bigg[2 F_D\left(\sqrt{\frac{k_B T}{k_a}}\right)
\\&\hspace{3cm}+ F_D\left(2 \sqrt{\frac{k_B T}{k_a}}\right)  \Bigg]
\end{align}
for $k_a\gg k_B T$. From these moments, we obtain the mean 
\begin{equation}
\bar{l}_0 = m_1/m_0 \simeq L_0(1 - k_B T/2k_a)
\end{equation}
and the standard deviation 
\begin{align}
\xi_\text{RL} &= \sqrt{m_2/m_0 - (m_1/m_0)^2} \\
&\simeq \sqrt{(k_B T/k_\text{RL}) + (k_BT L_0/2k_a)^2}
\end{align}
of the functions $\Omega_\text{RL}(l)$ and $K_\text{2D}(l)$ for $k_a\gg k_B T$. The mean value $\bar{l}_0$ is the preferred average separation of the membranes for large relative membrane roughnesses $\xi_\perp\gg \xi_\text{RL}$. 

From Eq. (\ref{moC}) and the rotational phase space volume
\begin{align}
\Omega_\text{R}  = \Omega_\text{L} & \simeq  2\pi \int_0^{\infty} e^{-\frac{1}{2} k_a \theta_a^2/k_B T} \sin \theta_a \, \text{d}\theta_a  \\
& \simeq \pi \sqrt{\frac{8 k_B T}{k_a}} F_D\left(\sqrt{\frac{k_B T}{2k_a}}\right)
\end{align}
of the unbound receptors and ligands, we obtain the integral
\begin{align}
\int K_\text{2D}(l) \text{d}l &\simeq \frac{K_\text{3D} k_a F_D\left(\sqrt{\frac{k_B T}{k_a}}\right)}{2\sqrt{k_a k_\text{RL}}\,\xi_z  F_D\left(\sqrt{\frac{k_B T}{2k_a}}\right)^2} 
\label{intK2D}\\
& \simeq \frac{K_\text{3D} k_a}{\sqrt{k_BT k_\text{RL}}\,\xi_z}
\label{intK2Dapprox}
\end{align}
of the function $K_\text{2D}(l)$. Eq.\ (\ref{intK2Dapprox}) results from the approximation $F_D(x)\simeq x$ for $x\ll 1$ of the Dawson function $F_D$ and is rather precise compared to Eq.\ (\ref{intK2D}), with a relative error of 0.1 \% for $k_a = 4\, k_B T$, and much smaller relative errors for larger values of $k_a$. Eq.\ (\ref{K2Dlim}) for the binding constant $K_\text{2D}$ at large membrane roughnesses $\xi_\perp\gg \xi_\text{RL}$ follows from Eq.\ (\ref{intK2Dapprox}).

\section{Roughness and variations of membrane normal of fluctuating membranes}

To obtain general scaling relations for the roughness and local orientation of fluctuating membranes, we consider here a tensionless quadratic membrane segment with projected area $L\times L$ and periodic boundary conditions in Monge parametrization. The shapes of this quadratic membrane segment can be described by the Fourier decomposition 
\begin{equation}
l(\boldsymbol{r}) = \sum_{\boldsymbol{q}} \left[a_{\boldsymbol{q}} \cos(\boldsymbol{q}\cdot \boldsymbol{r}) + b_{\boldsymbol{q}} \sin(\boldsymbol{q}\cdot \boldsymbol{r})\right]
\label{fourier}
\end{equation}
with $\boldsymbol{r} = (x,y)$ and $\boldsymbol{q} = (q_x, q_y) = 2 \pi (m,n)/L$ where $m$, and $n$ are integers. The summation in Eq.\ (\ref{fourier}) extends over half the $\boldsymbol{q}$-plane with $m\ge 0$. The bending energy of a given membrane shape with Fourier coefficients $\{a_{\boldsymbol{q}}\}$ and $\{b_{\boldsymbol{q}}\}$  then is
\begin{equation}
G = \int \frac{\kappa}{2} \left(\Delta l \right)^2 \text{d}x \,\text{d}y = \sum_{\boldsymbol{q}} \frac{\kappa}{4} q^4\left(a_{\boldsymbol{q}}^2 + b_{\boldsymbol{q}}^2\right)L^2
\label{bending_energy_fourier}
\end{equation}
with $q = \sqrt{q_x^2+q_y^2}$. Since the Fourier modes are decoupled, the mean-squared amplitude of each mode can be determined independently as
\begin{equation}
\left\langle a_{\boldsymbol{q}}^2 \right\rangle = \left\langle b_{\boldsymbol{q}}^2 \right\rangle 
= \frac{\int_{-\infty}^{\infty} b_{\boldsymbol{q}}^2 e^{-\kappa q^4 b_{\boldsymbol{q}}^2L^2/ 4 k_B T} \text{d}b_{\boldsymbol{q}}}{\int_{-\infty}^{\infty} e^{-\kappa q^4 b_{\boldsymbol{q}}^2L^2/ 4 k_B T} \text{d}b_{\boldsymbol{q}}} =
\frac{2 k_B T}{\kappa q^4 L^2}
\end{equation}

\begin{figure}[t]
\resizebox{\columnwidth}{!}{\includegraphics{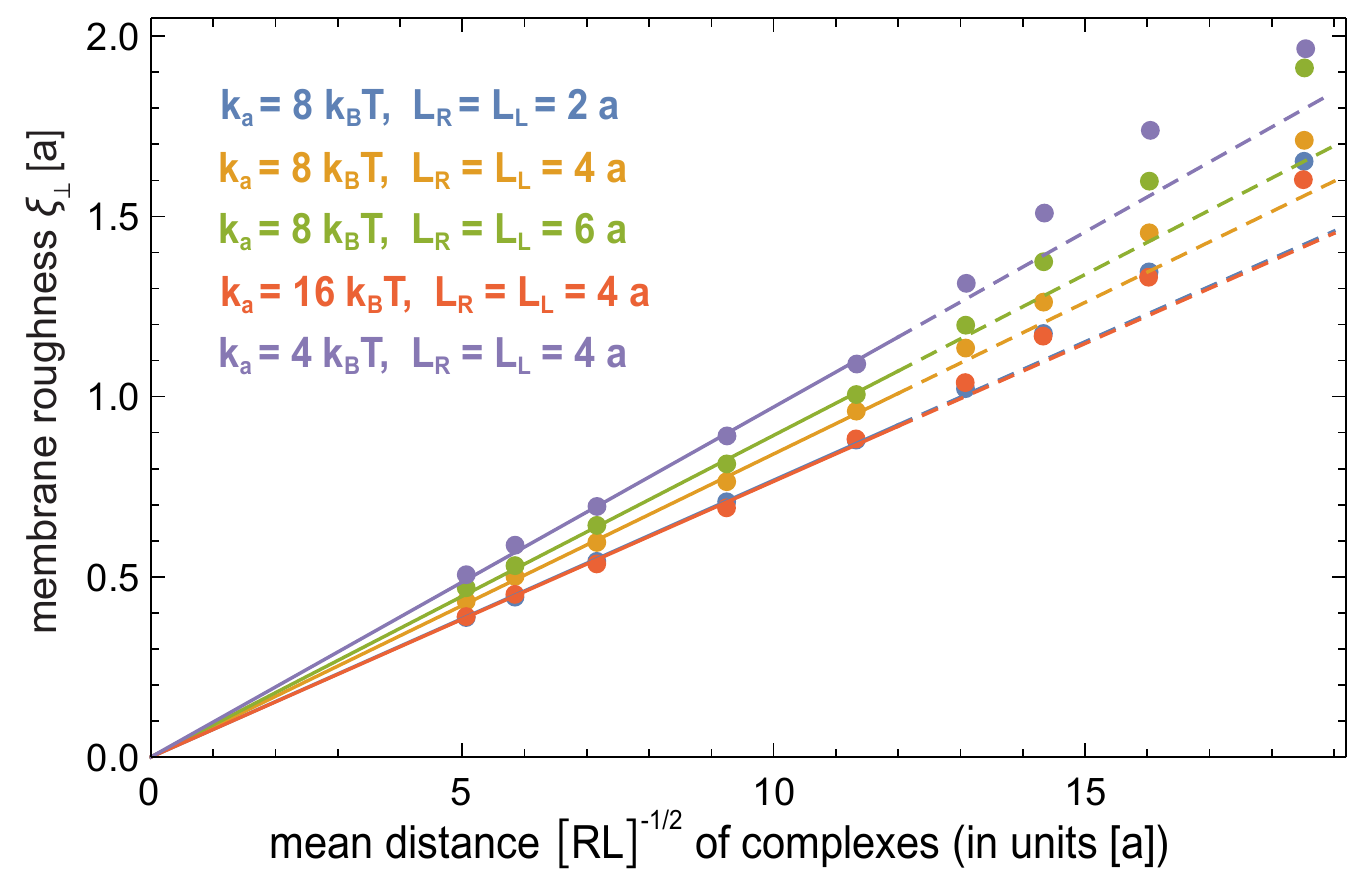}}
%\resizebox{0.6\columnwidth}{!}{\includegraphics{figure9}}
\caption{Relative roughness $\xi_\perp$ of two adhering, tensionless membranes versus the mean distance $1/\sqrt{[\text{RL}]}$ of neighboring RL complexes of our MC simulations. The relative roughness is proportional to $1/\sqrt{[\text{RL}]}$ in the range $\xi_\perp < a$, in agreement with the scaling relation (\ref{relative_roughness_scaling}). The values for the numerical prefactor $c_\perp$ of Eq.\ (\ref{relative_roughness_scaling}) obtained from linear fits in this roughness range (see full lines) are $c_\perp\simeq 0.17$ for the lengths $L_R= L_L = 2\, a$ and anchoring strength $k_a = 8\, k_B T$ of receptors and ligands, 
$c_\perp\simeq 0.19$ for $L_R= L_L = 4\, a$ and $k_a = 8\, k_B T$,  $c_\perp\simeq 0.20$ for $L_R= L_L = 6\, a$ and $k_a = 8\, k_B T$,  $c_\perp\simeq 0.17$ for $L_R= L_L = 4\, a$ and $k_a = 16\, k_B T$, and $c_\perp\simeq 0.22$ for $L_R= L_L = 4\, a$ and $k_a = 4\, k_B T$. 
}
\label{roughness_scaling}
\end{figure}

The local mean-square deviation of the membrane from the average location $\left\langle l(\boldsymbol{r}) \right \rangle = 0$ then can be calculated
\begin{align}
\left\langle l(\boldsymbol{r})^2\right \rangle 
&=  \sum_{\boldsymbol{q}} \left[\left\langle a_{\boldsymbol{q}}^2 \right\rangle \cos^2(\boldsymbol{q}\cdot \boldsymbol{r}) + \left\langle b_{\boldsymbol{q}}^2\right\rangle \sin^2(\boldsymbol{q}\cdot \boldsymbol{r}) \right]  \nonumber\\
&= \sum_{\boldsymbol{q}}\frac{2 k_B T}{\kappa q^4 L^2} \nonumber\\
&\simeq \left(\frac{L}{2\pi}\right)^2\int_{\pi/L}^{\pi/a}\frac{2 k_B T}{\kappa q^4 L^2} \pi q\,\text{d}q
\simeq \frac{k_B TL^2}{4\pi^3 \kappa} 
\label{msd}
\end{align}
after converting the sum over the wavevectors $\boldsymbol{q}$ into an integral over half the $\boldsymbol{q}$-plane from $q_\text{min}\simeq \pi/L$ to $q_\text{max}\simeq \pi/a$ where $a\ll L$ is molecular length scale. Similarly, the local mean-square gradient of the on average planar membrane can be calculated as
\begin{align}
 \left\langle \left( \nabla l(\boldsymbol{r})\right)^2\right \rangle & =  \sum_{\boldsymbol{q}} \left[q^2\left\langle a_{\boldsymbol{q}}^2 \right\rangle \sin^2(\boldsymbol{q}\cdot \boldsymbol{r}) + q^2\left\langle b_{\boldsymbol{q}}^2\right\rangle \cos^2(\boldsymbol{q}\cdot \boldsymbol{r}) \right]\nonumber \\
&= \sum_{\boldsymbol{q}}\frac{2 k_B T}{\kappa q^2 L^2} \nonumber\\
&\simeq \left(\frac{L}{2\pi}\right)^2\int_{\pi/L}^{\pi/a}\frac{2 k_B T}{\kappa q^2 L^2} \pi q\,\text{d}q = \frac{k_B T}{2\kappa\pi} \ln\left(\frac{L}{a}\right)
\label{mgrads}
\end{align}
According to Eq.\ (\ref{msd}), the roughness $\xi_\perp = \sqrt{\left\langle l(\boldsymbol{r})^2\right\rangle}$ is proportional to the linear size $L$ of the quadratic membrane segment, which in turn is proportional to the lateral correlation length $\xi_\parallel$ of the membrane. In our MC simulations with tensionless membranes, the lateral correlation length is proportional to the mean distance $1/\sqrt{[\text{RL}]}$ of neighboring RL complexes. Since the fluctuations of the separation field $l = l_1 - l_2$ of the two apposing membranes are governed by a bending energy of the form of Eq.~(\ref{bending_energy_fourier}) with effective bending rigidity $\kappa_\text{eff} = \kappa_1\kappa_2/( \kappa_1 + \kappa_2)$ where $\kappa_1$ and $\kappa_2$ are the rigidities of the two membranes \cite{Lipowsky88}, we obtain the scaling relation
\begin{equation}
\xi_\perp \simeq c_\perp \sqrt{k_B T / \kappa_\text{eff}} \left(\sqrt{[\text{RL}]}\right)^{-1/2}
\label{relative_roughness_scaling}
\end{equation}
between the relative roughness $\xi_\perp$ of the apposing membranes and the concentration $[\text{RL}]$ of receptor-ligand complexes.
Fig.\ \ref{roughness_scaling} illustrates that the relative roughness $\xi_\perp$ in our tensionless MC simulations is proportional to $1/\sqrt{[\text{RL}]}$ in the roughness range $\xi_\perp< a \simeq 5$ nm, in accordance with the scaling relation (\ref{relative_roughness_scaling}).
Linear fits in this roughness range lead to values of the numerical prefactor $c_\perp$ between 0.17 and 0.22, slightly depending on the length and anchoring strength of the receptor and ligand molecules. For larger relative roughnesses  $\xi_\perp>a$ , the fluctuation-mediated repulsion of the membranes leads to deviations from this linear scaling (see also Fig.\ \ref{figure_MC-fluc-L4}(b)). The effective rigidity of the two apposing membranes in our MC simulations with rigidities $\kappa_1 = \kappa_2 = 10\, k_BT$ is  $\kappa_\text{eff} = 5\, k_B T$. Previous MC simulations with receptor-ligand bonds that strongly constrain the local separation $l$ led to the value $c_\perp\simeq 0.14$ for the numerical prefactor  in Eq.\ (\ref{relative_roughness_scaling}). In general, the numerical prefactor $c_\perp$ depends on how strongly the receptor-ligand bonds constrain the membrane fluctuations, which can be quantified by the effective width $\xi_\text{RL}$ of the function $K_\text{2D}(l)$ given in Eq.\ (\ref{xiRL}).

\section{Effect of orientational variations of membrane normals on receptor-ligand binding.}

\begin{figure}[t]
\resizebox{\columnwidth}{!}{\includegraphics{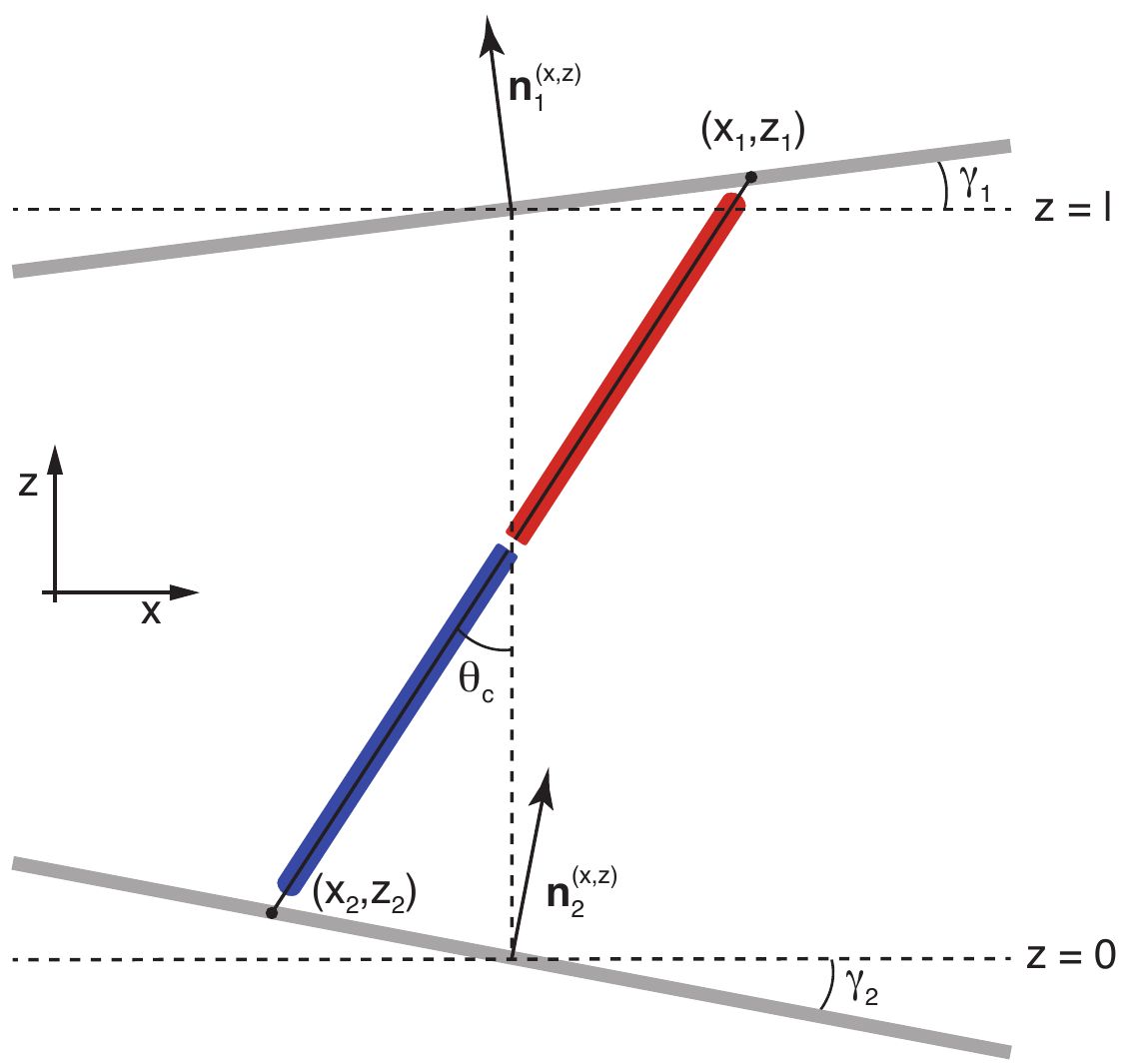}}
%\resizebox{0.6\columnwidth}{!}{\includegraphics{figure10}}
\caption{A bound receptor-ligand complex anchored to two membranes with local tilt angles $\gamma_1$ and $\gamma_2$ in the $x$-$z$ plane in which the complex is located. Here, $\boldsymbol{n}_1^{(x,z)}$ and $\boldsymbol{n}_2^{(x,z)}$ are the projections of the local normal vectors $\boldsymbol{n}_1$ and  $\boldsymbol{n}_2$ of the two membranes into the $x$-$z$ plane.
}
\label{membrane_tilt}
\end{figure}

Membrane shape fluctuations lead to orientational variations of the membrane normals. In this section, we consider how such variations affect the rotational phase space volume $\Omega_\text{RL}$ and, thus, the binding constant $K_\text{2D}$ of the receptor-ligand complexes. We focus on a single RL complex. The normals of the two membranes at the $(x,y)$ position of the center of mass of this complex are:
\begin{align}
\boldsymbol{n}_1 & = \left(\sin \theta_1 \cos \phi_1, \sin \theta_1 \sin \phi_1, \cos \theta_1\right) \\
\boldsymbol{n}_2 & = \left(\sin \theta_2 \cos \phi_2, \sin \theta_2 \sin \phi_2, \cos \theta_2\right)
\end{align}
Without loss of generality, we assume that the receptor-ligand complex is tilted in the $x$-$z$ plane (see Fig.\ \ref{membrane_tilt}). The orientation of the complex  then can be described by the unit vector
\begin{equation}
\boldsymbol{r}_c =  \left(\sin\theta_c, 0, \cos\theta_c\right)
\end{equation}
where $\theta_c$ is the tilt angle of the complex. The anchoring angles of the RL complex in the two membranes, i.e.\ the angles between $\boldsymbol{r}_c$ and the normals $\boldsymbol{n}_1$ and $\boldsymbol{n}_2$, then are
\begin{align}
\theta_{a,1} &= \arccos\left(\sin \theta_1 \cos \phi_1\sin\theta_c+ \cos \theta_1\cos\theta_c\right)
\label{thetaa1}\\
\theta_{a,2} &= \arccos\left(\sin \theta_2 \cos \phi_2\sin\theta_c+ \cos \theta_2\cos\theta_c\right)
\label{thetaa2}
\end{align} 
The tilt angles $\gamma_1$ and $\gamma_2$ of the projections of the normal vectors $\boldsymbol{n}_1$ and  $\boldsymbol{n}_2$ into the $x$-$z$ plane  (see Fig.\ \ref{membrane_tilt}) fulfill the relations
\begin{align}
\tan \gamma_1 &= \sin \theta_1 \cos \phi_1/\cos \theta_1 \\
\tan \gamma_2 &= \sin \theta_2 \cos \phi_2/\cos \theta_2 
\end{align}
The position of the anchor in membrane 1 then has the coordinates
\begin{align}
x_1 & = (L_\text{RL}/2) \sin \theta_c \\
z_1 & = l - x_1 \tan\ \gamma_1 = l - L_\text{RL} \sin \theta_c \sin \theta_1 \cos \phi_1/(2 \cos \theta_1)
\end{align}
where $L_\text{RL}$ is the length of the rod-like RL complex, and $l$ is the local deviation of membrane 1 at the center of mass of the complex (see Fig.\ \ref{membrane_tilt}). The position of the anchor in membrane 2 has the coordinates
\begin{align}
x_2 & = - (L_\text{RL}/2) \sin \theta_c \\
z_2 & = - x_2 \tan \gamma_2 = - L_\text{RL} \sin \theta_c \sin \theta_2 \cos \phi_2/(2 \cos \theta_2)
\end{align}
From the relation $L_\text{RL}^2 = (x_1 - x_2)^2 + (z_1 - z_2)^2$ between the length $L_\text{RL}$ of the complex and positions of its membrane anchors, we obtain 
\begin{equation}
L_\text{RL} = 2 l /\left( 2\cos \theta_c + \sin\theta_c (\tan \theta_1 \cos \phi_1 - \tan\theta_2 \cos \phi_2)\right)
\label{LRLgen}
\end{equation}
This equation reduces to Eq.\ (16) of planar membranes for $\theta_1 = \theta_2 = 0$.

The effective configurational energy (\ref{Hef}) now can be generalized to 
\begin{equation}
H_\text{RL}(\theta_c,l,\boldsymbol{n}_1,\boldsymbol{n}_2)  \simeq  \frac{1}{2} k_a \left(\theta_{a,1}^2 +  \theta_{a,2}^2 \right) 
+ \frac{1}{2} k_\text{RL} \left(L_\text{RL}- L_o\right)^2
\end{equation}
with $\theta_{a,1}$, $\theta_{a,2}$, and $L_\text{RL}$ given in Eqs.\ (\ref{thetaa1}),  (\ref{thetaa2}), and (\ref{LRLgen}). The rotational phase space volume of the RL complex for a fixed orientation of the normals and fixed local separation $l$ at the center of mass of the complex is then
\begin{equation}
\Omega_\text{RL}(l,\boldsymbol{n}_1,\boldsymbol{n}_2) 
\simeq 2 \pi \int_0^{\pi/2} e^{-H_\text{RL}(\theta_c,l,\boldsymbol{n}_1,\boldsymbol{n}_2)/k_B T}  \sin\theta_c \text{d}\theta_c
\label{OmegaRLgen}
\end{equation}

\begin{figure}[t]
\resizebox{\columnwidth}{!}{\includegraphics{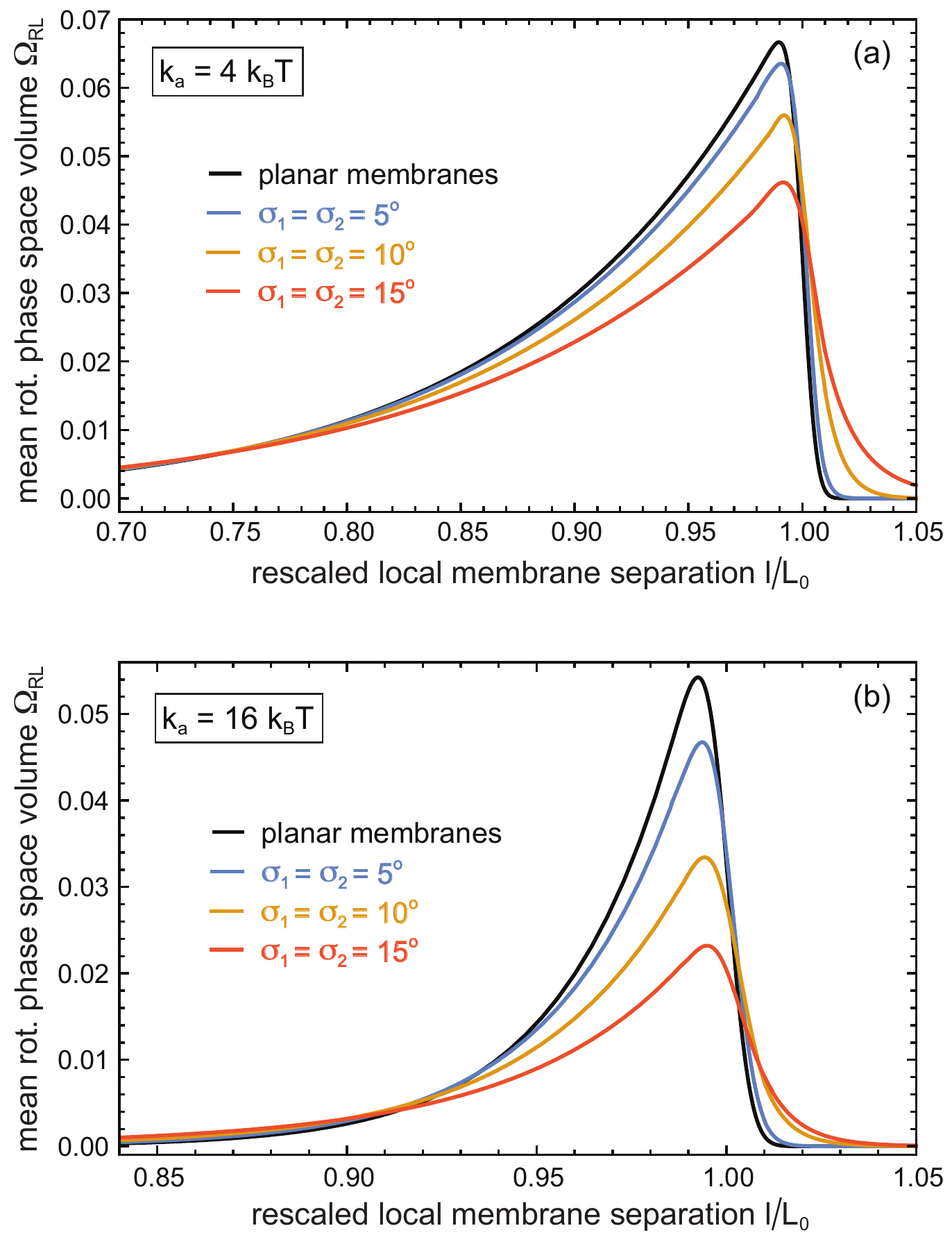}}
%\resizebox{0.6\columnwidth}{!}{\includegraphics{figure11}}
\caption{Mean rotational phase volume $\Omega_\text{RL}$ as a function of the rescaled local separation $l/L_0$ for orientational fluctuations of the membrane normals with standard deviations $\sigma_1 = \sigma_2$. The anchoring strength of the receptors and ligands is $k_a = 4\, k_B T$ in (a) and  $k_a = 16\, k_B T$ in (b), and the effective stiffness of the receptor-ligand complex has the value $k_\text{RL} L_0^2 = 4.5 \cdot 10^4 \,k_B T$.  The mean rotational phase space volume is calculated as average of $\Omega_\text{RL}$ for 10000 orientations of the normal vectors $\boldsymbol{n}_1$ and  $\boldsymbol{n}_2$ that are randomly chosen from Gaussian distributions with standard deviations $\sigma_1$ and  $\sigma_2$.}
\label{OmegaRL_with_tilt}
\end{figure}

Fig.\ \ref{OmegaRL_with_tilt} illustrates how fluctuations of the membrane normals affect the function $\Omega_\text{RL}(l)$, for Gaussian distributions of the tilt angles $\theta_1$ and $\theta_2$ of the normal vectors with various standard deviations $\sigma_1 = \sigma_2$. For fixed local separation $l$, the values of  $\Omega_\text{RL}$ shown in Fig.\ \ref{OmegaRL_with_tilt} are averages over 10000 randomly  chosen orientations of the normal vectors $\boldsymbol{n}_1$ and  $\boldsymbol{n}_2$ with standard deviations  $\sigma_1$ and $\sigma_2$ of the tilt angles $\theta_1$ and $\theta_2$. The maximum value of the function $\Omega_\text{RL}(l)$ decreases for increasing standard deviations $\sigma_1 = \sigma_2$ of the tilt angles of the normal vectors, while the width of this function increases. These changes of the function  $\Omega_\text{RL}(l)$ due to fluctuations of the normal vectors are more pronounced for the larger anchoring strength $k_a = 16\, k_B T$ of the receptors and ligands, compared to $k_a = 4\, k_B T$. In general, the effect of fluctuations of the normal vectors on $\Omega_\text{RL}(l)$ and $K_\text{2D}(l)$ can be expected to be small if the standard deviations $\sigma_1$ and $\sigma_2$ of the tilt angles of the normal vectors are small compared to the standard deviation of the anchoring angles, which increase with decreasing anchoring strength $k_a$.

In our MC simulations with tensionless membranes, the standard deviations of the orientational variations of the membrane normals are about $\sigma_1 = \sigma_2 \simeq 9.8^\circ$ for the relative membrane roughness $\xi_\perp\simeq 0.5\,a $ and $\sigma_1 = \sigma_2 \simeq 11^\circ$ for $\xi_\perp\simeq 1.0\,a$. These values have been obtained from MC simulations with the anchoring strength $k_a = 8\, k_B T$ and lengths $L_R = L_L = 4\,a$ of the receptors and ligands. From Eq.\ (\ref{msd}) with effective rigidity $\kappa_\text{eff}= \kappa/2$ relevant for the relative roughness of two apposing membranes with equal rigidities $\kappa_1 = \kappa_2 = \kappa$ (see Appendix D) and from Eq.\ (\ref{mgrads}), we obtain
\begin{equation}
 \left\langle \left( \nabla l(\boldsymbol{r})\right)^2\right \rangle \simeq \frac{k_B T}{2\kappa\pi}\ln\left(\frac{\sqrt{2}\pi^{3/2}\xi_\perp}{a \sqrt{k_B T/ \kappa}}\right)
 \label{Eplus}
\end{equation}
for the local mean-square tilt of the membranes. Eq.\ (\ref{Eplus})  leads to the estimates $\sigma_1 = \sigma_2 \simeq 11.3^\circ$ and $12.7^\circ$ for the relative roughnesses  $\xi_\perp\simeq 0.5\,a $ and $\xi_\perp\simeq 1.0\,a$, respectively, and the membrane rigidity $\kappa= 10\, k_B T$ of our MC simulations, which are slightly larger than the measured values of $\sigma_1$ and $\sigma_2$ given above. 

For relative membrane roughnesses $\xi_\perp$ that are large compared to the width $\xi_\text{RL}$ of the functions $\Omega_\text{RL}(l)$ and $K_\text{2D}(l)$, the binding constant $K_\text{2D}$ only depends on the integrals of these functions (see Eq.\ (21)). In Fig.\ \ref{OmegaRL_with_tilt}(a), the integral of the function $\Omega_\text{RL}(l)$ is reduced by 
$1\%$, $6\%$, and $12\%$ for $\sigma_1 = \sigma_2 = 5^\circ$, $10^\circ$, and $15^\circ$, respectively, compared to the integral of $\Omega_\text{RL}(l)$ for planar membranes.
In Fig.\ \ref{OmegaRL_with_tilt}(b), the integral of $\Omega_\text{RL}(l)$ is reduced by 
$6\%$, $20\%$, and $34\%$ for $\sigma_1 = \sigma_2 = 5^\circ$, $10^\circ$, and $15^\circ$, respectively, compared to the integral of $\Omega_\text{RL}(l)$ for planar membranes. 
For the relative membrane roughness  $\xi_\perp\simeq 1.0\,a$ with $\sigma_1 = \sigma_2 \simeq 11^\circ$ obtained from our MC simulations (see above), the fluctuations of the membrane normals thus effectively reduce the $K_\text{2D}$ values of our theory by about 7\% for the anchoring strength $k_a = 4 \, k_B T$, and by about 23\% for the anchoring strength $k_a =16 \, k_B T$. For the anchoring strengths $k_a = 16\, k_B T$, the full and the dashed theory lines in Fig.\ 4 indeed overestimate the $K_\text{2D}$ data points from MC simulations with fluctuating membranes by about 13\% and 8\%, respectively, for large roughnesses around  $\xi_\perp\simeq 1.0\,a$, which is somewhat smaller than the estimate of  23\% above obtained from taking into account the fluctuations of the normals. The relative error of our MC data points is about 5\%.  

% If in two-column mode, this environment will change to single-column format so that long equations can be displayed. 
%\begin{widetext}
%$$\mbox{put long equation here}$$
%\end{widetext}

% Figures should be put into the text as floats. 
% Use the graphics or graphicx packages (distributed with LaTeX2e).
% Put the label that you will use with \ref{} command in the braces of the \label{} command.
%
% \begin{figure}
% \end{figure}

% \begin{table}
% \end{tabular}
% \end{table}

%\begin{acknowledgments}
% Put your acknowledgments here.
%\end{acknowledgments}

%\bibliography{membranes}

\begin{thebibliography}{69}%
\makeatletter
\providecommand \@ifxundefined [1]{%
 \@ifx{#1\undefined}
}%
\providecommand \@ifnum [1]{%
 \ifnum #1\expandafter \@firstoftwo
 \else \expandafter \@secondoftwo
 \fi
}%
\providecommand \@ifx [1]{%
 \ifx #1\expandafter \@firstoftwo
 \else \expandafter \@secondoftwo
 \fi
}%
\providecommand \natexlab [1]{#1}%
\providecommand \enquote  [1]{``#1''}%
\providecommand \bibnamefont  [1]{#1}%
\providecommand \bibfnamefont [1]{#1}%
\providecommand \citenamefont [1]{#1}%
\providecommand \href@noop [0]{\@secondoftwo}%
\providecommand \href [0]{\begingroup \@sanitize@url \@href}%
\providecommand \@href[1]{\@@startlink{#1}\@@href}%
\providecommand \@@href[1]{\endgroup#1\@@endlink}%
\providecommand \@sanitize@url [0]{\catcode `\\12\catcode `\$12\catcode
  `\&12\catcode `\#12\catcode `\^12\catcode `\_12\catcode `\%12\relax}%
\providecommand \@@startlink[1]{}%
\providecommand \@@endlink[0]{}%
\providecommand \url  [0]{\begingroup\@sanitize@url \@url }%
\providecommand \@url [1]{\endgroup\@href {#1}{\urlprefix }}%
\providecommand \urlprefix  [0]{URL }%
\providecommand \Eprint [0]{\href }%
\providecommand \doibase [0]{http://dx.doi.org/}%
\providecommand \selectlanguage [0]{\@gobble}%
\providecommand \bibinfo  [0]{\@secondoftwo}%
\providecommand \bibfield  [0]{\@secondoftwo}%
\providecommand \translation [1]{[#1]}%
\providecommand \BibitemOpen [0]{}%
\providecommand \bibitemStop [0]{}%
\providecommand \bibitemNoStop [0]{.\EOS\space}%
\providecommand \EOS [0]{\spacefactor3000\relax}%
\providecommand \BibitemShut  [1]{\csname bibitem#1\endcsname}%
\let\auto@bib@innerbib\@empty
%</preamble>
\bibitem [{\citenamefont {Dustin}\ \emph {et~al.}(2001)\citenamefont {Dustin},
  \citenamefont {Bromley}, \citenamefont {Davis},\ and\ \citenamefont
  {Zhu}}]{Dustin01}%
  \BibitemOpen
  \bibfield  {author} {\bibinfo {author} {\bibfnamefont {M.~L.}\ \bibnamefont
  {Dustin}}, \bibinfo {author} {\bibfnamefont {S.~K.}\ \bibnamefont {Bromley}},
  \bibinfo {author} {\bibfnamefont {M.~M.}\ \bibnamefont {Davis}}, \ and\
  \bibinfo {author} {\bibfnamefont {C.}~\bibnamefont {Zhu}},\ }\href {\doibase
  10.1146/annurev.cellbio.17.1.133} {\bibfield  {journal} {\bibinfo  {journal}
  {Annu. Rev. Cell Dev. Biol.}\ }\textbf {\bibinfo {volume} {17}},\ \bibinfo
  {pages} {133} (\bibinfo {year} {2001})}\BibitemShut {NoStop}%
\bibitem [{\citenamefont {Orsello}, \citenamefont {Lauffenburger},\ and\
  \citenamefont {Hammer}(2001)}]{Orsello01}%
  \BibitemOpen
  \bibfield  {author} {\bibinfo {author} {\bibfnamefont {C.~E.}\ \bibnamefont
  {Orsello}}, \bibinfo {author} {\bibfnamefont {D.~A.}\ \bibnamefont
  {Lauffenburger}}, \ and\ \bibinfo {author} {\bibfnamefont {D.~A.}\
  \bibnamefont {Hammer}},\ }\href@noop {} {\bibfield  {journal} {\bibinfo
  {journal} {Trends Biotechnol.}\ }\textbf {\bibinfo {volume} {19}},\ \bibinfo
  {pages} {310} (\bibinfo {year} {2001})}\BibitemShut {NoStop}%
\bibitem [{\citenamefont {Leckband}\ and\ \citenamefont
  {Sivasankar}(2012)}]{Leckband12}%
  \BibitemOpen
  \bibfield  {author} {\bibinfo {author} {\bibfnamefont {D.}~\bibnamefont
  {Leckband}}\ and\ \bibinfo {author} {\bibfnamefont {S.}~\bibnamefont
  {Sivasankar}},\ }\href {\doibase 10.1016/j.ceb.2012.05.014} {\bibfield
  {journal} {\bibinfo  {journal} {Curr. Opin. Cell. Biol.}\ }\textbf {\bibinfo
  {volume} {24}},\ \bibinfo {pages} {620} (\bibinfo {year} {2012})}\BibitemShut
  {NoStop}%
\bibitem [{\citenamefont {Zarnitsyna}\ and\ \citenamefont
  {Zhu}(2012)}]{Zarnitsyna12}%
  \BibitemOpen
  \bibfield  {author} {\bibinfo {author} {\bibfnamefont {V.}~\bibnamefont
  {Zarnitsyna}}\ and\ \bibinfo {author} {\bibfnamefont {C.}~\bibnamefont
  {Zhu}},\ }\href {\doibase 10.1088/1478-3975/9/4/045005} {\bibfield  {journal}
  {\bibinfo  {journal} {Phys. Biol.}\ }\textbf {\bibinfo {volume} {9}},\
  \bibinfo {pages} {045005} (\bibinfo {year} {2012})}\BibitemShut {NoStop}%
\bibitem [{\citenamefont {Krobath}\ \emph {et~al.}(2009)\citenamefont
  {Krobath}, \citenamefont {Rozycki}, \citenamefont {Lipowsky},\ and\
  \citenamefont {Weikl}}]{Krobath09}%
  \BibitemOpen
  \bibfield  {author} {\bibinfo {author} {\bibfnamefont {H.}~\bibnamefont
  {Krobath}}, \bibinfo {author} {\bibfnamefont {B.}~\bibnamefont {Rozycki}},
  \bibinfo {author} {\bibfnamefont {R.}~\bibnamefont {Lipowsky}}, \ and\
  \bibinfo {author} {\bibfnamefont {T.~R.}\ \bibnamefont {Weikl}},\ }\href@noop
  {} {\bibfield  {journal} {\bibinfo  {journal} {Soft Matter}\ }\textbf
  {\bibinfo {volume} {5}},\ \bibinfo {pages} {3354} (\bibinfo {year}
  {2009})}\BibitemShut {NoStop}%
\bibitem [{\citenamefont {Wu}\ \emph {et~al.}(2010)\citenamefont {Wu},
  \citenamefont {Jin}, \citenamefont {Harrison}, \citenamefont {Shapiro},
  \citenamefont {Honig},\ and\ \citenamefont {Ben-Shaul}}]{Wu10}%
  \BibitemOpen
  \bibfield  {author} {\bibinfo {author} {\bibfnamefont {Y.}~\bibnamefont
  {Wu}}, \bibinfo {author} {\bibfnamefont {X.}~\bibnamefont {Jin}}, \bibinfo
  {author} {\bibfnamefont {O.}~\bibnamefont {Harrison}}, \bibinfo {author}
  {\bibfnamefont {L.}~\bibnamefont {Shapiro}}, \bibinfo {author} {\bibfnamefont
  {B.~H.}\ \bibnamefont {Honig}}, \ and\ \bibinfo {author} {\bibfnamefont
  {A.}~\bibnamefont {Ben-Shaul}},\ }\href {\doibase 10.1073/pnas.1011247107}
  {\bibfield  {journal} {\bibinfo  {journal} {Proc. Natl. Acad. Sci. USA}\
  }\textbf {\bibinfo {volume} {107}},\ \bibinfo {pages} {17592} (\bibinfo
  {year} {2010})}\BibitemShut {NoStop}%
\bibitem [{\citenamefont {Hu}, \citenamefont {Lipowsky},\ and\ \citenamefont
  {Weikl}(2013)}]{Hu13}%
  \BibitemOpen
  \bibfield  {author} {\bibinfo {author} {\bibfnamefont {J.}~\bibnamefont
  {Hu}}, \bibinfo {author} {\bibfnamefont {R.}~\bibnamefont {Lipowsky}}, \ and\
  \bibinfo {author} {\bibfnamefont {T.~R.}\ \bibnamefont {Weikl}},\ }\href@noop
  {} {\bibfield  {journal} {\bibinfo  {journal} {Proc. Natl. Acad. Sci. USA}\
  }\textbf {\bibinfo {volume} {110}},\ \bibinfo {pages} {15283} (\bibinfo
  {year} {2013})}\BibitemShut {NoStop}%
\bibitem [{\citenamefont {Wu}, \citenamefont {Honig},\ and\ \citenamefont
  {Ben-Shaul}(2013)}]{Wu13}%
  \BibitemOpen
  \bibfield  {author} {\bibinfo {author} {\bibfnamefont {Y.}~\bibnamefont
  {Wu}}, \bibinfo {author} {\bibfnamefont {B.}~\bibnamefont {Honig}}, \ and\
  \bibinfo {author} {\bibfnamefont {A.}~\bibnamefont {Ben-Shaul}},\ }\href
  {\doibase 10.1016/j.bpj.2013.02.009} {\bibfield  {journal} {\bibinfo
  {journal} {Biophys. J.}\ }\textbf {\bibinfo {volume} {104}},\ \bibinfo
  {pages} {1221} (\bibinfo {year} {2013})}\BibitemShut {NoStop}%
\bibitem [{\citenamefont {Schuck}(1997)}]{Schuck97}%
  \BibitemOpen
  \bibfield  {author} {\bibinfo {author} {\bibfnamefont {P.}~\bibnamefont
  {Schuck}},\ }\href {\doibase 10.1146/annurev.biophys.26.1.541} {\bibfield
  {journal} {\bibinfo  {journal} {Annu. Rev. Biophys. Biomol. Struct.}\
  }\textbf {\bibinfo {volume} {26}},\ \bibinfo {pages} {541} (\bibinfo {year}
  {1997})}\BibitemShut {NoStop}%
\bibitem [{\citenamefont {Rich}\ and\ \citenamefont {Myszka}(2000)}]{Rich00}%
  \BibitemOpen
  \bibfield  {author} {\bibinfo {author} {\bibfnamefont {R.~L.}\ \bibnamefont
  {Rich}}\ and\ \bibinfo {author} {\bibfnamefont {D.~G.}\ \bibnamefont
  {Myszka}},\ }\href@noop {} {\bibfield  {journal} {\bibinfo  {journal} {Curr.
  Opin. Biotechnol.}\ }\textbf {\bibinfo {volume} {11}},\ \bibinfo {pages} {54}
  (\bibinfo {year} {2000})}\BibitemShut {NoStop}%
\bibitem [{\citenamefont {McDonnell}(2001)}]{McDonnell01}%
  \BibitemOpen
  \bibfield  {author} {\bibinfo {author} {\bibfnamefont {J.~M.}\ \bibnamefont
  {McDonnell}},\ }\href@noop {} {\bibfield  {journal} {\bibinfo  {journal}
  {Curr. Opin. Chem. Biol.}\ }\textbf {\bibinfo {volume} {5}},\ \bibinfo
  {pages} {572} (\bibinfo {year} {2001})}\BibitemShut {NoStop}%
\bibitem [{\citenamefont {Dustin}\ \emph {et~al.}(1996)\citenamefont {Dustin},
  \citenamefont {Ferguson}, \citenamefont {Chan}, \citenamefont {Springer},\
  and\ \citenamefont {Golan}}]{Dustin96}%
  \BibitemOpen
  \bibfield  {author} {\bibinfo {author} {\bibfnamefont {M.~L.}\ \bibnamefont
  {Dustin}}, \bibinfo {author} {\bibfnamefont {L.~M.}\ \bibnamefont
  {Ferguson}}, \bibinfo {author} {\bibfnamefont {P.~Y.}\ \bibnamefont {Chan}},
  \bibinfo {author} {\bibfnamefont {T.~A.}\ \bibnamefont {Springer}}, \ and\
  \bibinfo {author} {\bibfnamefont {D.~E.}\ \bibnamefont {Golan}},\ }\href@noop
  {} {\bibfield  {journal} {\bibinfo  {journal} {J. Cell. Biol.}\ }\textbf
  {\bibinfo {volume} {132}},\ \bibinfo {pages} {465} (\bibinfo {year}
  {1996})}\BibitemShut {NoStop}%
\bibitem [{\citenamefont {Dustin}\ \emph {et~al.}(1997)\citenamefont {Dustin},
  \citenamefont {Golan}, \citenamefont {Zhu}, \citenamefont {Miller},
  \citenamefont {Meier}, \citenamefont {Davies},\ and\ \citenamefont {van~der
  Merwe}}]{Dustin97}%
  \BibitemOpen
  \bibfield  {author} {\bibinfo {author} {\bibfnamefont {M.~L.}\ \bibnamefont
  {Dustin}}, \bibinfo {author} {\bibfnamefont {D.~E.}\ \bibnamefont {Golan}},
  \bibinfo {author} {\bibfnamefont {D.~M.}\ \bibnamefont {Zhu}}, \bibinfo
  {author} {\bibfnamefont {J.~M.}\ \bibnamefont {Miller}}, \bibinfo {author}
  {\bibfnamefont {W.}~\bibnamefont {Meier}}, \bibinfo {author} {\bibfnamefont
  {E.~A.}\ \bibnamefont {Davies}}, \ and\ \bibinfo {author} {\bibfnamefont
  {P.~A.}\ \bibnamefont {van~der Merwe}},\ }\href@noop {} {\bibfield  {journal}
  {\bibinfo  {journal} {J. Biol. Chem.}\ }\textbf {\bibinfo {volume} {272}},\
  \bibinfo {pages} {30889} (\bibinfo {year} {1997})}\BibitemShut {NoStop}%
\bibitem [{\citenamefont {Zhu}\ \emph {et~al.}(2007)\citenamefont {Zhu},
  \citenamefont {Dustin}, \citenamefont {Cairo},\ and\ \citenamefont
  {Golan}}]{Zhu07}%
  \BibitemOpen
  \bibfield  {author} {\bibinfo {author} {\bibfnamefont {D.-M.}\ \bibnamefont
  {Zhu}}, \bibinfo {author} {\bibfnamefont {M.~L.}\ \bibnamefont {Dustin}},
  \bibinfo {author} {\bibfnamefont {C.~W.}\ \bibnamefont {Cairo}}, \ and\
  \bibinfo {author} {\bibfnamefont {D.~E.}\ \bibnamefont {Golan}},\ }\href@noop
  {} {\bibfield  {journal} {\bibinfo  {journal} {Biophys. J.}\ }\textbf
  {\bibinfo {volume} {92}},\ \bibinfo {pages} {1022} (\bibinfo {year}
  {2007})}\BibitemShut {NoStop}%
\bibitem [{\citenamefont {Tolentino}\ \emph {et~al.}(2008)\citenamefont
  {Tolentino}, \citenamefont {Wu}, \citenamefont {Zarnitsyna}, \citenamefont
  {Fang}, \citenamefont {Dustin},\ and\ \citenamefont {Zhu}}]{Tolentino08}%
  \BibitemOpen
  \bibfield  {author} {\bibinfo {author} {\bibfnamefont {T.~P.}\ \bibnamefont
  {Tolentino}}, \bibinfo {author} {\bibfnamefont {J.}~\bibnamefont {Wu}},
  \bibinfo {author} {\bibfnamefont {V.~I.}\ \bibnamefont {Zarnitsyna}},
  \bibinfo {author} {\bibfnamefont {Y.}~\bibnamefont {Fang}}, \bibinfo {author}
  {\bibfnamefont {M.~L.}\ \bibnamefont {Dustin}}, \ and\ \bibinfo {author}
  {\bibfnamefont {C.}~\bibnamefont {Zhu}},\ }\href {\doibase
  10.1529/biophysj.107.114447} {\bibfield  {journal} {\bibinfo  {journal}
  {Biophys. J.}\ }\textbf {\bibinfo {volume} {95}},\ \bibinfo {pages} {920}
  (\bibinfo {year} {2008})}\BibitemShut {NoStop}%
\bibitem [{\citenamefont {Huppa}\ \emph {et~al.}(2010)\citenamefont {Huppa},
  \citenamefont {Axmann}, \citenamefont {M{\"o}rtelmaier}, \citenamefont
  {Lillemeier}, \citenamefont {Newell}, \citenamefont {Brameshuber},
  \citenamefont {Klein}, \citenamefont {Sch\"utz},\ and\ \citenamefont
  {Davis}}]{Huppa10}%
  \BibitemOpen
  \bibfield  {author} {\bibinfo {author} {\bibfnamefont {J.~B.}\ \bibnamefont
  {Huppa}}, \bibinfo {author} {\bibfnamefont {M.}~\bibnamefont {Axmann}},
  \bibinfo {author} {\bibfnamefont {M.~A.}\ \bibnamefont {M{\"o}rtelmaier}},
  \bibinfo {author} {\bibfnamefont {B.~F.}\ \bibnamefont {Lillemeier}},
  \bibinfo {author} {\bibfnamefont {E.~W.}\ \bibnamefont {Newell}}, \bibinfo
  {author} {\bibfnamefont {M.}~\bibnamefont {Brameshuber}}, \bibinfo {author}
  {\bibfnamefont {L.~O.}\ \bibnamefont {Klein}}, \bibinfo {author}
  {\bibfnamefont {G.~J.}\ \bibnamefont {Sch\"utz}}, \ and\ \bibinfo {author}
  {\bibfnamefont {M.~M.}\ \bibnamefont {Davis}},\ }\href {\doibase
  10.1038/nature08746} {\bibfield  {journal} {\bibinfo  {journal} {Nature}\
  }\textbf {\bibinfo {volume} {463}},\ \bibinfo {pages} {963} (\bibinfo {year}
  {2010})}\BibitemShut {NoStop}%
\bibitem [{\citenamefont {Axmann}\ \emph {et~al.}(2012)\citenamefont {Axmann},
  \citenamefont {Huppa}, \citenamefont {Davis},\ and\ \citenamefont
  {Sch\"utz}}]{Axmann12}%
  \BibitemOpen
  \bibfield  {author} {\bibinfo {author} {\bibfnamefont {M.}~\bibnamefont
  {Axmann}}, \bibinfo {author} {\bibfnamefont {J.~B.}\ \bibnamefont {Huppa}},
  \bibinfo {author} {\bibfnamefont {M.~M.}\ \bibnamefont {Davis}}, \ and\
  \bibinfo {author} {\bibfnamefont {G.~J.}\ \bibnamefont {Sch\"utz}},\
  }\href@noop {} {\bibfield  {journal} {\bibinfo  {journal} {Biophys. J.}\
  }\textbf {\bibinfo {volume} {103}},\ \bibinfo {pages} {L17} (\bibinfo {year}
  {2012})}\BibitemShut {NoStop}%
\bibitem [{\citenamefont {O'Donoghue}\ \emph {et~al.}(2013)\citenamefont
  {O'Donoghue}, \citenamefont {Pielak}, \citenamefont {Smoligovets},
  \citenamefont {Lin},\ and\ \citenamefont {Groves}}]{ODonoghue13}%
  \BibitemOpen
  \bibfield  {author} {\bibinfo {author} {\bibfnamefont {G.~P.}\ \bibnamefont
  {O'Donoghue}}, \bibinfo {author} {\bibfnamefont {R.~M.}\ \bibnamefont
  {Pielak}}, \bibinfo {author} {\bibfnamefont {A.~A.}\ \bibnamefont
  {Smoligovets}}, \bibinfo {author} {\bibfnamefont {J.~J.}\ \bibnamefont
  {Lin}}, \ and\ \bibinfo {author} {\bibfnamefont {J.~T.}\ \bibnamefont
  {Groves}},\ }\href {\doibase 10.7554/eLife.00778} {\bibfield  {journal}
  {\bibinfo  {journal} {Elife}\ }\textbf {\bibinfo {volume} {2}},\ \bibinfo
  {pages} {e00778} (\bibinfo {year} {2013})}\BibitemShut {NoStop}%
\bibitem [{\citenamefont {Kaplanski}\ \emph {et~al.}(1993)\citenamefont
  {Kaplanski}, \citenamefont {Farnarier}, \citenamefont {Tissot}, \citenamefont
  {Pierres}, \citenamefont {Benoliel}, \citenamefont {Alessi}, \citenamefont
  {Kaplanski},\ and\ \citenamefont {Bongrand}}]{Kaplanski93}%
  \BibitemOpen
  \bibfield  {author} {\bibinfo {author} {\bibfnamefont {G.}~\bibnamefont
  {Kaplanski}}, \bibinfo {author} {\bibfnamefont {C.}~\bibnamefont
  {Farnarier}}, \bibinfo {author} {\bibfnamefont {O.}~\bibnamefont {Tissot}},
  \bibinfo {author} {\bibfnamefont {A.}~\bibnamefont {Pierres}}, \bibinfo
  {author} {\bibfnamefont {A.~M.}\ \bibnamefont {Benoliel}}, \bibinfo {author}
  {\bibfnamefont {M.~C.}\ \bibnamefont {Alessi}}, \bibinfo {author}
  {\bibfnamefont {S.}~\bibnamefont {Kaplanski}}, \ and\ \bibinfo {author}
  {\bibfnamefont {P.}~\bibnamefont {Bongrand}},\ }\href@noop {} {\bibfield
  {journal} {\bibinfo  {journal} {Biophys. J.}\ }\textbf {\bibinfo {volume}
  {64}},\ \bibinfo {pages} {1922} (\bibinfo {year} {1993})}\BibitemShut
  {NoStop}%
\bibitem [{\citenamefont {Alon}, \citenamefont {Hammer},\ and\ \citenamefont
  {Springer}(1995)}]{Alon95}%
  \BibitemOpen
  \bibfield  {author} {\bibinfo {author} {\bibfnamefont {R.}~\bibnamefont
  {Alon}}, \bibinfo {author} {\bibfnamefont {D.~A.}\ \bibnamefont {Hammer}}, \
  and\ \bibinfo {author} {\bibfnamefont {T.~A.}\ \bibnamefont {Springer}},\
  }\href {\doibase 10.1038/374539a0} {\bibfield  {journal} {\bibinfo  {journal}
  {Nature}\ }\textbf {\bibinfo {volume} {374}},\ \bibinfo {pages} {539}
  (\bibinfo {year} {1995})}\BibitemShut {NoStop}%
\bibitem [{\citenamefont {Piper}, \citenamefont {Swerlick},\ and\ \citenamefont
  {Zhu}(1998)}]{Piper98}%
  \BibitemOpen
  \bibfield  {author} {\bibinfo {author} {\bibfnamefont {J.~W.}\ \bibnamefont
  {Piper}}, \bibinfo {author} {\bibfnamefont {R.~A.}\ \bibnamefont {Swerlick}},
  \ and\ \bibinfo {author} {\bibfnamefont {C.}~\bibnamefont {Zhu}},\ }\href
  {\doibase 10.1016/S0006-3495(98)77807-5} {\bibfield  {journal} {\bibinfo
  {journal} {Biophys. J.}\ }\textbf {\bibinfo {volume} {74}},\ \bibinfo {pages}
  {492} (\bibinfo {year} {1998})}\BibitemShut {NoStop}%
\bibitem [{\citenamefont {Chesla}, \citenamefont {Selvaraj},\ and\
  \citenamefont {Zhu}(1998)}]{Chesla98}%
  \BibitemOpen
  \bibfield  {author} {\bibinfo {author} {\bibfnamefont {S.~E.}\ \bibnamefont
  {Chesla}}, \bibinfo {author} {\bibfnamefont {P.}~\bibnamefont {Selvaraj}}, \
  and\ \bibinfo {author} {\bibfnamefont {C.}~\bibnamefont {Zhu}},\ }\href@noop
  {} {\bibfield  {journal} {\bibinfo  {journal} {Biophys. J.}\ }\textbf
  {\bibinfo {volume} {75}},\ \bibinfo {pages} {1553} (\bibinfo {year}
  {1998})}\BibitemShut {NoStop}%
\bibitem [{\citenamefont {Merkel}\ \emph {et~al.}(1999)\citenamefont {Merkel},
  \citenamefont {Nassoy}, \citenamefont {Leung}, \citenamefont {Ritchie},\ and\
  \citenamefont {Evans}}]{Merkel99}%
  \BibitemOpen
  \bibfield  {author} {\bibinfo {author} {\bibfnamefont {R.}~\bibnamefont
  {Merkel}}, \bibinfo {author} {\bibfnamefont {P.}~\bibnamefont {Nassoy}},
  \bibinfo {author} {\bibfnamefont {A.}~\bibnamefont {Leung}}, \bibinfo
  {author} {\bibfnamefont {K.}~\bibnamefont {Ritchie}}, \ and\ \bibinfo
  {author} {\bibfnamefont {E.}~\bibnamefont {Evans}},\ }\href {\doibase
  10.1038/16219} {\bibfield  {journal} {\bibinfo  {journal} {Nature}\ }\textbf
  {\bibinfo {volume} {397}},\ \bibinfo {pages} {50} (\bibinfo {year}
  {1999})}\BibitemShut {NoStop}%
\bibitem [{\citenamefont {Williams}\ \emph {et~al.}(2001)\citenamefont
  {Williams}, \citenamefont {Nagarajan}, \citenamefont {Selvaraj},\ and\
  \citenamefont {Zhu}}]{Williams01}%
  \BibitemOpen
  \bibfield  {author} {\bibinfo {author} {\bibfnamefont {T.~E.}\ \bibnamefont
  {Williams}}, \bibinfo {author} {\bibfnamefont {S.}~\bibnamefont {Nagarajan}},
  \bibinfo {author} {\bibfnamefont {P.}~\bibnamefont {Selvaraj}}, \ and\
  \bibinfo {author} {\bibfnamefont {C.}~\bibnamefont {Zhu}},\ }\href {\doibase
  10.1074/jbc.M010427200} {\bibfield  {journal} {\bibinfo  {journal} {J. Biol.
  Chem.}\ }\textbf {\bibinfo {volume} {276}},\ \bibinfo {pages} {13283}
  (\bibinfo {year} {2001})}\BibitemShut {NoStop}%
\bibitem [{\citenamefont {Chen}\ \emph {et~al.}(2008)\citenamefont {Chen},
  \citenamefont {Evans}, \citenamefont {McEver},\ and\ \citenamefont
  {Zhu}}]{Chen08}%
  \BibitemOpen
  \bibfield  {author} {\bibinfo {author} {\bibfnamefont {W.}~\bibnamefont
  {Chen}}, \bibinfo {author} {\bibfnamefont {E.~A.}\ \bibnamefont {Evans}},
  \bibinfo {author} {\bibfnamefont {R.~P.}\ \bibnamefont {McEver}}, \ and\
  \bibinfo {author} {\bibfnamefont {C.}~\bibnamefont {Zhu}},\ }\href {\doibase
  10.1529/biophysj.107.117895} {\bibfield  {journal} {\bibinfo  {journal}
  {Biophys. J.}\ }\textbf {\bibinfo {volume} {94}},\ \bibinfo {pages} {694}
  (\bibinfo {year} {2008})}\BibitemShut {NoStop}%
\bibitem [{\citenamefont {Huang}\ \emph {et~al.}(2010)\citenamefont {Huang},
  \citenamefont {Zarnitsyna}, \citenamefont {Liu}, \citenamefont {Edwards},
  \citenamefont {Jiang}, \citenamefont {Evavold},\ and\ \citenamefont
  {Zhu}}]{Huang10}%
  \BibitemOpen
  \bibfield  {author} {\bibinfo {author} {\bibfnamefont {J.}~\bibnamefont
  {Huang}}, \bibinfo {author} {\bibfnamefont {V.~I.}\ \bibnamefont
  {Zarnitsyna}}, \bibinfo {author} {\bibfnamefont {B.}~\bibnamefont {Liu}},
  \bibinfo {author} {\bibfnamefont {L.~J.}\ \bibnamefont {Edwards}}, \bibinfo
  {author} {\bibfnamefont {N.}~\bibnamefont {Jiang}}, \bibinfo {author}
  {\bibfnamefont {B.~D.}\ \bibnamefont {Evavold}}, \ and\ \bibinfo {author}
  {\bibfnamefont {C.}~\bibnamefont {Zhu}},\ }\href@noop {} {\bibfield
  {journal} {\bibinfo  {journal} {Nature}\ }\textbf {\bibinfo {volume} {464}},\
  \bibinfo {pages} {932} (\bibinfo {year} {2010})}\BibitemShut {NoStop}%
\bibitem [{\citenamefont {Liu}\ \emph {et~al.}(2014)\citenamefont {Liu},
  \citenamefont {Chen}, \citenamefont {Evavold},\ and\ \citenamefont
  {Zhu}}]{Liu14}%
  \BibitemOpen
  \bibfield  {author} {\bibinfo {author} {\bibfnamefont {B.}~\bibnamefont
  {Liu}}, \bibinfo {author} {\bibfnamefont {W.}~\bibnamefont {Chen}}, \bibinfo
  {author} {\bibfnamefont {B.~D.}\ \bibnamefont {Evavold}}, \ and\ \bibinfo
  {author} {\bibfnamefont {C.}~\bibnamefont {Zhu}},\ }\href {\doibase
  10.1016/j.cell.2014.02.053} {\bibfield  {journal} {\bibinfo  {journal}
  {Cell}\ }\textbf {\bibinfo {volume} {157}},\ \bibinfo {pages} {357} (\bibinfo
  {year} {2014})}\BibitemShut {NoStop}%
\bibitem [{\citenamefont {Lipowsky}(1996)}]{Lipowsky96}%
  \BibitemOpen
  \bibfield  {author} {\bibinfo {author} {\bibfnamefont {R.}~\bibnamefont
  {Lipowsky}},\ }\href@noop {} {\bibfield  {journal} {\bibinfo  {journal}
  {Phys. Rev. Lett.}\ }\textbf {\bibinfo {volume} {77}},\ \bibinfo {pages}
  {1652} (\bibinfo {year} {1996})}\BibitemShut {NoStop}%
\bibitem [{\citenamefont {Weikl}\ and\ \citenamefont
  {Lipowsky}(2001)}]{Weikl01}%
  \BibitemOpen
  \bibfield  {author} {\bibinfo {author} {\bibfnamefont {T.~R.}\ \bibnamefont
  {Weikl}}\ and\ \bibinfo {author} {\bibfnamefont {R.}~\bibnamefont
  {Lipowsky}},\ }\href@noop {} {\bibfield  {journal} {\bibinfo  {journal}
  {Phys. Rev. E.}\ }\textbf {\bibinfo {volume} {64}},\ \bibinfo {pages}
  {011903} (\bibinfo {year} {2001})}\BibitemShut {NoStop}%
\bibitem [{\citenamefont {Weikl}, \citenamefont {Groves},\ and\ \citenamefont
  {Lipowsky}(2002)}]{Weikl02a}%
  \BibitemOpen
  \bibfield  {author} {\bibinfo {author} {\bibfnamefont {T.~R.}\ \bibnamefont
  {Weikl}}, \bibinfo {author} {\bibfnamefont {J.~T.}\ \bibnamefont {Groves}}, \
  and\ \bibinfo {author} {\bibfnamefont {R.}~\bibnamefont {Lipowsky}},\
  }\href@noop {} {\bibfield  {journal} {\bibinfo  {journal} {Europhys. Lett.}\
  }\textbf {\bibinfo {volume} {59}},\ \bibinfo {pages} {916} (\bibinfo {year}
  {2002})}\BibitemShut {NoStop}%
\bibitem [{\citenamefont {Weikl}\ and\ \citenamefont
  {Lipowsky}(2004)}]{Weikl04}%
  \BibitemOpen
  \bibfield  {author} {\bibinfo {author} {\bibfnamefont {T.~R.}\ \bibnamefont
  {Weikl}}\ and\ \bibinfo {author} {\bibfnamefont {R.}~\bibnamefont
  {Lipowsky}},\ }\href@noop {} {\bibfield  {journal} {\bibinfo  {journal}
  {Biophys. J.}\ }\textbf {\bibinfo {volume} {87}},\ \bibinfo {pages} {3665}
  (\bibinfo {year} {2004})}\BibitemShut {NoStop}%
\bibitem [{\citenamefont {Asfaw}\ \emph {et~al.}(2006)\citenamefont {Asfaw},
  \citenamefont {Rozycki}, \citenamefont {Lipowsky},\ and\ \citenamefont
  {Weikl}}]{Asfaw06}%
  \BibitemOpen
  \bibfield  {author} {\bibinfo {author} {\bibfnamefont {M.}~\bibnamefont
  {Asfaw}}, \bibinfo {author} {\bibfnamefont {B.}~\bibnamefont {Rozycki}},
  \bibinfo {author} {\bibfnamefont {R.}~\bibnamefont {Lipowsky}}, \ and\
  \bibinfo {author} {\bibfnamefont {T.~R.}\ \bibnamefont {Weikl}},\ }\href@noop
  {} {\bibfield  {journal} {\bibinfo  {journal} {Europhys. Lett.}\ }\textbf
  {\bibinfo {volume} {76}},\ \bibinfo {pages} {703} (\bibinfo {year}
  {2006})}\BibitemShut {NoStop}%
\bibitem [{\citenamefont {Tsourkas}\ \emph {et~al.}(2007)\citenamefont
  {Tsourkas}, \citenamefont {Baumgarth}, \citenamefont {Simon},\ and\
  \citenamefont {Raychaudhuri}}]{Tsourkas07}%
  \BibitemOpen
  \bibfield  {author} {\bibinfo {author} {\bibfnamefont {P.~K.}\ \bibnamefont
  {Tsourkas}}, \bibinfo {author} {\bibfnamefont {N.}~\bibnamefont {Baumgarth}},
  \bibinfo {author} {\bibfnamefont {S.~I.}\ \bibnamefont {Simon}}, \ and\
  \bibinfo {author} {\bibfnamefont {S.}~\bibnamefont {Raychaudhuri}},\ }\href
  {\doibase 10.1529/biophysj.106.094995} {\bibfield  {journal} {\bibinfo
  {journal} {Biophys. J.}\ }\textbf {\bibinfo {volume} {92}},\ \bibinfo {pages}
  {4196} (\bibinfo {year} {2007})}\BibitemShut {NoStop}%
\bibitem [{\citenamefont {Reister-Gottfried}\ \emph {et~al.}(2008)\citenamefont
  {Reister-Gottfried}, \citenamefont {Sengupta}, \citenamefont {Lorz},
  \citenamefont {Sackmann}, \citenamefont {Seifert},\ and\ \citenamefont
  {Smith}}]{Reister08}%
  \BibitemOpen
  \bibfield  {author} {\bibinfo {author} {\bibfnamefont {E.}~\bibnamefont
  {Reister-Gottfried}}, \bibinfo {author} {\bibfnamefont {K.}~\bibnamefont
  {Sengupta}}, \bibinfo {author} {\bibfnamefont {B.}~\bibnamefont {Lorz}},
  \bibinfo {author} {\bibfnamefont {E.}~\bibnamefont {Sackmann}}, \bibinfo
  {author} {\bibfnamefont {U.}~\bibnamefont {Seifert}}, \ and\ \bibinfo
  {author} {\bibfnamefont {A.~S.}\ \bibnamefont {Smith}},\ }\href@noop {}
  {\bibfield  {journal} {\bibinfo  {journal} {Phys. Rev. Lett.}\ }\textbf
  {\bibinfo {volume} {101}},\ \bibinfo {pages} {208103} (\bibinfo {year}
  {2008})}\BibitemShut {NoStop}%
\bibitem [{\citenamefont {Bihr}, \citenamefont {Seifert},\ and\ \citenamefont
  {Smith}(2012)}]{Bihr12}%
  \BibitemOpen
  \bibfield  {author} {\bibinfo {author} {\bibfnamefont {T.}~\bibnamefont
  {Bihr}}, \bibinfo {author} {\bibfnamefont {U.}~\bibnamefont {Seifert}}, \
  and\ \bibinfo {author} {\bibfnamefont {A.-S.}\ \bibnamefont {Smith}},\
  }\href@noop {} {\bibfield  {journal} {\bibinfo  {journal} {Phys. Rev. Lett.}\
  }\textbf {\bibinfo {volume} {109}},\ \bibinfo {pages} {258101} (\bibinfo
  {year} {2012})}\BibitemShut {NoStop}%
\bibitem [{\citenamefont {Komura}\ and\ \citenamefont
  {Andelman}(2000)}]{Komura00}%
  \BibitemOpen
  \bibfield  {author} {\bibinfo {author} {\bibfnamefont {S.}~\bibnamefont
  {Komura}}\ and\ \bibinfo {author} {\bibfnamefont {D.}~\bibnamefont
  {Andelman}},\ }\href@noop {} {\bibfield  {journal} {\bibinfo  {journal} {Eur.
  Phys. J. E}\ }\textbf {\bibinfo {volume} {3}},\ \bibinfo {pages} {259}
  (\bibinfo {year} {2000})}\BibitemShut {NoStop}%
\bibitem [{\citenamefont {Bruinsma}, \citenamefont {Behrisch},\ and\
  \citenamefont {Sackmann}(2000)}]{Bruinsma00}%
  \BibitemOpen
  \bibfield  {author} {\bibinfo {author} {\bibfnamefont {R.}~\bibnamefont
  {Bruinsma}}, \bibinfo {author} {\bibfnamefont {A.}~\bibnamefont {Behrisch}},
  \ and\ \bibinfo {author} {\bibfnamefont {E.}~\bibnamefont {Sackmann}},\
  }\href@noop {} {\bibfield  {journal} {\bibinfo  {journal} {Phys. Rev. E}\
  }\textbf {\bibinfo {volume} {61}},\ \bibinfo {pages} {4253} (\bibinfo {year}
  {2000})}\BibitemShut {NoStop}%
\bibitem [{\citenamefont {Qi}, \citenamefont {Groves},\ and\ \citenamefont
  {Chakraborty}(2001)}]{Qi01}%
  \BibitemOpen
  \bibfield  {author} {\bibinfo {author} {\bibfnamefont {S.~Y.}\ \bibnamefont
  {Qi}}, \bibinfo {author} {\bibfnamefont {J.~T.}\ \bibnamefont {Groves}}, \
  and\ \bibinfo {author} {\bibfnamefont {A.~K.}\ \bibnamefont {Chakraborty}},\
  }\href {\doibase 10.1073/pnas.111536798} {\bibfield  {journal} {\bibinfo
  {journal} {Proc. Natl. Acad. Sci. USA}\ }\textbf {\bibinfo {volume} {98}},\
  \bibinfo {pages} {6548} (\bibinfo {year} {2001})}\BibitemShut {NoStop}%
\bibitem [{\citenamefont {Chen}(2003)}]{Chen03}%
  \BibitemOpen
  \bibfield  {author} {\bibinfo {author} {\bibfnamefont {H.-Y.}\ \bibnamefont
  {Chen}},\ }\href@noop {} {\bibfield  {journal} {\bibinfo  {journal} {Phys.
  Rev. E}\ }\textbf {\bibinfo {volume} {67}},\ \bibinfo {pages} {031919}
  (\bibinfo {year} {2003})}\BibitemShut {NoStop}%
\bibitem [{\citenamefont {Raychaudhuri}, \citenamefont {Chakraborty},\ and\
  \citenamefont {Kardar}(2003)}]{Raychaudhuri03}%
  \BibitemOpen
  \bibfield  {author} {\bibinfo {author} {\bibfnamefont {S.}~\bibnamefont
  {Raychaudhuri}}, \bibinfo {author} {\bibfnamefont {A.~K.}\ \bibnamefont
  {Chakraborty}}, \ and\ \bibinfo {author} {\bibfnamefont {M.}~\bibnamefont
  {Kardar}},\ }\href@noop {} {\bibfield  {journal} {\bibinfo  {journal} {Phys.
  Rev. Lett.}\ }\textbf {\bibinfo {volume} {91}},\ \bibinfo {pages} {208101}
  (\bibinfo {year} {2003})}\BibitemShut {NoStop}%
\bibitem [{\citenamefont {Coombs}\ \emph {et~al.}(2004)\citenamefont {Coombs},
  \citenamefont {Dembo}, \citenamefont {Wofsy},\ and\ \citenamefont
  {Goldstein}}]{Coombs04}%
  \BibitemOpen
  \bibfield  {author} {\bibinfo {author} {\bibfnamefont {D.}~\bibnamefont
  {Coombs}}, \bibinfo {author} {\bibfnamefont {M.}~\bibnamefont {Dembo}},
  \bibinfo {author} {\bibfnamefont {C.}~\bibnamefont {Wofsy}}, \ and\ \bibinfo
  {author} {\bibfnamefont {B.}~\bibnamefont {Goldstein}},\ }\href@noop {}
  {\bibfield  {journal} {\bibinfo  {journal} {Biophys. J.}\ }\textbf {\bibinfo
  {volume} {86}},\ \bibinfo {pages} {1408} (\bibinfo {year}
  {2004})}\BibitemShut {NoStop}%
\bibitem [{\citenamefont {Shenoy}\ and\ \citenamefont
  {Freund}(2005)}]{Shenoy05}%
  \BibitemOpen
  \bibfield  {author} {\bibinfo {author} {\bibfnamefont {V.~B.}\ \bibnamefont
  {Shenoy}}\ and\ \bibinfo {author} {\bibfnamefont {L.~B.}\ \bibnamefont
  {Freund}},\ }\href {\doibase 10.1073/pnas.0500368102} {\bibfield  {journal}
  {\bibinfo  {journal} {Proc. Natl. Acad. Sci. USA}\ }\textbf {\bibinfo
  {volume} {102}},\ \bibinfo {pages} {3213} (\bibinfo {year}
  {2005})}\BibitemShut {NoStop}%
\bibitem [{\citenamefont {Wu}\ and\ \citenamefont {Chen}(2006)}]{Wu06}%
  \BibitemOpen
  \bibfield  {author} {\bibinfo {author} {\bibfnamefont {J.-Y.}\ \bibnamefont
  {Wu}}\ and\ \bibinfo {author} {\bibfnamefont {H.-Y.}\ \bibnamefont {Chen}},\
  }\href@noop {} {\bibfield  {journal} {\bibinfo  {journal} {Phys. Rev. E}\
  }\textbf {\bibinfo {volume} {73}},\ \bibinfo {pages} {011914} (\bibinfo
  {year} {2006})}\BibitemShut {NoStop}%
\bibitem [{\citenamefont {Zuckerman}\ and\ \citenamefont
  {Bruinsma}(1995)}]{Zuckerman95}%
  \BibitemOpen
  \bibfield  {author} {\bibinfo {author} {\bibfnamefont {D.}~\bibnamefont
  {Zuckerman}}\ and\ \bibinfo {author} {\bibfnamefont {R.}~\bibnamefont
  {Bruinsma}},\ }\href@noop {} {\bibfield  {journal} {\bibinfo  {journal}
  {Phys. Rev. Lett.}\ }\textbf {\bibinfo {volume} {74}},\ \bibinfo {pages}
  {3900} (\bibinfo {year} {1995})}\BibitemShut {NoStop}%
\bibitem [{\citenamefont {Krobath}\ \emph {et~al.}(2007)\citenamefont
  {Krobath}, \citenamefont {Sch\"utz}, \citenamefont {Lipowsky},\ and\
  \citenamefont {Weikl}}]{Krobath07}%
  \BibitemOpen
  \bibfield  {author} {\bibinfo {author} {\bibfnamefont {H.}~\bibnamefont
  {Krobath}}, \bibinfo {author} {\bibfnamefont {G.~J.}\ \bibnamefont
  {Sch\"utz}}, \bibinfo {author} {\bibfnamefont {R.}~\bibnamefont {Lipowsky}},
  \ and\ \bibinfo {author} {\bibfnamefont {T.~R.}\ \bibnamefont {Weikl}},\
  }\href@noop {} {\bibfield  {journal} {\bibinfo  {journal} {Europhys. Lett.}\
  }\textbf {\bibinfo {volume} {78}},\ \bibinfo {pages} {38003} (\bibinfo {year}
  {2007})}\BibitemShut {NoStop}%
\bibitem [{\citenamefont {Hu}\ \emph {et~al.}()\citenamefont {Hu},
  \citenamefont {Xu}, \citenamefont {Lipowsky},\ and\ \citenamefont
  {Weikl}}]{Hu}%
  \BibitemOpen
  \bibfield  {author} {\bibinfo {author} {\bibfnamefont {J.}~\bibnamefont
  {Hu}}, \bibinfo {author} {\bibfnamefont {G.-K.}\ \bibnamefont {Xu}}, \bibinfo
  {author} {\bibfnamefont {R.}~\bibnamefont {Lipowsky}}, \ and\ \bibinfo
  {author} {\bibfnamefont {T.~R.}\ \bibnamefont {Weikl}},\ }\href@noop {}
  {\bibinfo  {journal} {Accompanying manuscript}\ }\BibitemShut {NoStop}%
\bibitem [{\citenamefont {Lipowsky}\ and\ \citenamefont
  {Zielinska}(1989)}]{Lipowsky89}%
  \BibitemOpen
\bibfield  {journal} {  }\bibfield  {author} {\bibinfo {author} {\bibnamefont
  {Lipowsky}}\ and\ \bibinfo {author} {\bibnamefont {Zielinska}},\ }\href@noop
  {} {\bibfield  {journal} {\bibinfo  {journal} {Phys. Rev. Lett.}\ }\textbf
  {\bibinfo {volume} {62}},\ \bibinfo {pages} {1572} (\bibinfo {year}
  {1989})}\BibitemShut {NoStop}%
\bibitem [{\citenamefont {Weikl}\ and\ \citenamefont
  {Lipowsky}(2006)}]{Weikl06}%
  \BibitemOpen
  \bibfield  {author} {\bibinfo {author} {\bibfnamefont {T.~R.}\ \bibnamefont
  {Weikl}}\ and\ \bibinfo {author} {\bibfnamefont {R.}~\bibnamefont
  {Lipowsky}},\ }\href@noop {} {\emph {\bibinfo {title} {{Membrane adhesion and
  domain formation. {\it In} Advances in Planar Lipid Bilayers and Liposomes.
  A. Leitmannova Liu, editor}}}}\ (\bibinfo  {publisher} {Academic Press},\
  \bibinfo {year} {2006})\BibitemShut {NoStop}%
\bibitem [{\citenamefont {Helfrich}(1973)}]{Helfrich73}%
  \BibitemOpen
  \bibfield  {author} {\bibinfo {author} {\bibfnamefont {W.}~\bibnamefont
  {Helfrich}},\ }\href@noop {} {\bibfield  {journal} {\bibinfo  {journal} {Z.
  Naturforsch. C}\ }\textbf {\bibinfo {volume} {28}},\ \bibinfo {pages} {693}
  (\bibinfo {year} {1973})}\BibitemShut {NoStop}%
\bibitem [{\citenamefont {Goetz}, \citenamefont {Gompper},\ and\ \citenamefont
  {Lipowsky}(1999)}]{Goetz99}%
  \BibitemOpen
  \bibfield  {author} {\bibinfo {author} {\bibfnamefont {R.}~\bibnamefont
  {Goetz}}, \bibinfo {author} {\bibfnamefont {G.}~\bibnamefont {Gompper}}, \
  and\ \bibinfo {author} {\bibfnamefont {R.}~\bibnamefont {Lipowsky}},\
  }\href@noop {} {\bibfield  {journal} {\bibinfo  {journal} {Phys. Rev. Lett.}\
  }\textbf {\bibinfo {volume} {82}},\ \bibinfo {pages} {221} (\bibinfo {year}
  {1999})}\BibitemShut {NoStop}%
\bibitem [{\citenamefont {Luo}\ and\ \citenamefont {Sharp}(2002)}]{Luo02}%
  \BibitemOpen
  \bibfield  {author} {\bibinfo {author} {\bibfnamefont {H.}~\bibnamefont
  {Luo}}\ and\ \bibinfo {author} {\bibfnamefont {K.}~\bibnamefont {Sharp}},\
  }\href@noop {} {\bibfield  {journal} {\bibinfo  {journal} {Proc. Natl. Acad.
  Sci. USA}\ }\textbf {\bibinfo {volume} {99}},\ \bibinfo {pages} {10399}
  (\bibinfo {year} {2002})}\BibitemShut {NoStop}%
\bibitem [{\citenamefont {Woo}\ and\ \citenamefont {Roux}(2005)}]{Woo05}%
  \BibitemOpen
  \bibfield  {author} {\bibinfo {author} {\bibfnamefont {H.-J.}\ \bibnamefont
  {Woo}}\ and\ \bibinfo {author} {\bibfnamefont {B.}~\bibnamefont {Roux}},\
  }\href@noop {} {\bibfield  {journal} {\bibinfo  {journal} {Proc. Natl. Acad.
  Sci. USA}\ }\textbf {\bibinfo {volume} {102}},\ \bibinfo {pages} {6825}
  (\bibinfo {year} {2005})}\BibitemShut {NoStop}%
\bibitem [{Note1()}]{Note1}%
  \BibitemOpen
  \bibinfo {note} {The effect of the tilt of the receptor-ligand complexes
  relative to the membrane normal on $V_b/A_b$ can be taken into account via
  $\xi _x$ and $\xi _y$. However, since the values of the standard deviations
  $\xi _x$, $\xi _y$, and $\xi _z$ in the directions $x$ and $y$ perpendicular
  to the complex and the direction $z$ parallel to the complex are typically
  rather similar, we neglect this effect here.}\BibitemShut {Stop}%
\bibitem [{Note2()}]{Note2}%
  \BibitemOpen
  \bibinfo {note} {In contrast, related averages over local membrane
  separations for the on-rate constant $k_\protect \text {on}$ and off-rate
  constant $k_\protect \text {off}$ rely on characteristic timescales for
  membrane shape fluctuations that are much smaller than the characteristic
  timescales for the diffusion of the anchored molecules on the relevant length
  scales, and much smaller than the characteristic binding times \cite
  {Bihr12,Hu}}\BibitemShut {NoStop}%
\bibitem [{\citenamefont {Dimova}(2014)}]{Dimova14}%
  \BibitemOpen
  \bibfield  {author} {\bibinfo {author} {\bibfnamefont {R.}~\bibnamefont
  {Dimova}},\ }\href {\doibase 10.1016/j.cis.2014.03.003} {\bibfield  {journal}
  {\bibinfo  {journal} {Adv. Colloid Interface Sci.}\ }\textbf {\bibinfo
  {volume} {208}},\ \bibinfo {pages} {225} (\bibinfo {year}
  {2014})}\BibitemShut {NoStop}%
\bibitem [{\citenamefont {Monks}\ \emph {et~al.}(1998)\citenamefont {Monks},
  \citenamefont {Freiberg}, \citenamefont {Kupfer}, \citenamefont {Sciaky},\
  and\ \citenamefont {Kupfer}}]{Monks98}%
  \BibitemOpen
  \bibfield  {author} {\bibinfo {author} {\bibfnamefont {C.~R.}\ \bibnamefont
  {Monks}}, \bibinfo {author} {\bibfnamefont {B.~A.}\ \bibnamefont {Freiberg}},
  \bibinfo {author} {\bibfnamefont {H.}~\bibnamefont {Kupfer}}, \bibinfo
  {author} {\bibfnamefont {N.}~\bibnamefont {Sciaky}}, \ and\ \bibinfo {author}
  {\bibfnamefont {A.}~\bibnamefont {Kupfer}},\ }\href {\doibase 10.1038/25764}
  {\bibfield  {journal} {\bibinfo  {journal} {Nature}\ }\textbf {\bibinfo
  {volume} {395}},\ \bibinfo {pages} {82} (\bibinfo {year} {1998})}\BibitemShut
  {NoStop}%
\bibitem [{\citenamefont {Grakoui}\ \emph {et~al.}(1999)\citenamefont
  {Grakoui}, \citenamefont {Bromley}, \citenamefont {Sumen}, \citenamefont
  {Davis}, \citenamefont {Shaw}, \citenamefont {Allen},\ and\ \citenamefont
  {Dustin}}]{Grakoui99}%
  \BibitemOpen
  \bibfield  {author} {\bibinfo {author} {\bibfnamefont {A.}~\bibnamefont
  {Grakoui}}, \bibinfo {author} {\bibfnamefont {S.~K.}\ \bibnamefont
  {Bromley}}, \bibinfo {author} {\bibfnamefont {C.}~\bibnamefont {Sumen}},
  \bibinfo {author} {\bibfnamefont {M.~M.}\ \bibnamefont {Davis}}, \bibinfo
  {author} {\bibfnamefont {A.~S.}\ \bibnamefont {Shaw}}, \bibinfo {author}
  {\bibfnamefont {P.~M.}\ \bibnamefont {Allen}}, \ and\ \bibinfo {author}
  {\bibfnamefont {M.~L.}\ \bibnamefont {Dustin}},\ }\href@noop {} {\bibfield
  {journal} {\bibinfo  {journal} {Science}\ }\textbf {\bibinfo {volume}
  {285}},\ \bibinfo {pages} {221} (\bibinfo {year} {1999})}\BibitemShut
  {NoStop}%
\bibitem [{\citenamefont {Mossman}\ \emph {et~al.}(2005)\citenamefont
  {Mossman}, \citenamefont {Campi}, \citenamefont {Groves},\ and\ \citenamefont
  {Dustin}}]{Mossman05}%
  \BibitemOpen
  \bibfield  {author} {\bibinfo {author} {\bibfnamefont {K.~D.}\ \bibnamefont
  {Mossman}}, \bibinfo {author} {\bibfnamefont {G.}~\bibnamefont {Campi}},
  \bibinfo {author} {\bibfnamefont {J.~T.}\ \bibnamefont {Groves}}, \ and\
  \bibinfo {author} {\bibfnamefont {M.~L.}\ \bibnamefont {Dustin}},\ }\href
  {\doibase 10.1126/science.1119238} {\bibfield  {journal} {\bibinfo  {journal}
  {Science}\ }\textbf {\bibinfo {volume} {310}},\ \bibinfo {pages} {1191}
  (\bibinfo {year} {2005})}\BibitemShut {NoStop}%
\bibitem [{\citenamefont {Paszek}\ \emph {et~al.}(2014)\citenamefont {Paszek},
  \citenamefont {DuFort}, \citenamefont {Rossier}, \citenamefont {Bainer},
  \citenamefont {Mouw}, \citenamefont {Godula}, \citenamefont {Hudak},
  \citenamefont {Lakins}, \citenamefont {Wijekoon}, \citenamefont {Cassereau},
  \citenamefont {Rubashkin}, \citenamefont {Magbanua}, \citenamefont {Thorn},
  \citenamefont {Davidson}, \citenamefont {Rugo}, \citenamefont {Park},
  \citenamefont {Hammer}, \citenamefont {Giannone}, \citenamefont {Bertozzi},\
  and\ \citenamefont {Weaver}}]{Paszek14}%
  \BibitemOpen
  \bibfield  {author} {\bibinfo {author} {\bibfnamefont {M.~J.}\ \bibnamefont
  {Paszek}}, \bibinfo {author} {\bibfnamefont {C.~C.}\ \bibnamefont {DuFort}},
  \bibinfo {author} {\bibfnamefont {O.}~\bibnamefont {Rossier}}, \bibinfo
  {author} {\bibfnamefont {R.}~\bibnamefont {Bainer}}, \bibinfo {author}
  {\bibfnamefont {J.~K.}\ \bibnamefont {Mouw}}, \bibinfo {author}
  {\bibfnamefont {K.}~\bibnamefont {Godula}}, \bibinfo {author} {\bibfnamefont
  {J.~E.}\ \bibnamefont {Hudak}}, \bibinfo {author} {\bibfnamefont {J.~N.}\
  \bibnamefont {Lakins}}, \bibinfo {author} {\bibfnamefont {A.~C.}\
  \bibnamefont {Wijekoon}}, \bibinfo {author} {\bibfnamefont {L.}~\bibnamefont
  {Cassereau}}, \bibinfo {author} {\bibfnamefont {M.~G.}\ \bibnamefont
  {Rubashkin}}, \bibinfo {author} {\bibfnamefont {M.~J.}\ \bibnamefont
  {Magbanua}}, \bibinfo {author} {\bibfnamefont {K.~S.}\ \bibnamefont {Thorn}},
  \bibinfo {author} {\bibfnamefont {M.~W.}\ \bibnamefont {Davidson}}, \bibinfo
  {author} {\bibfnamefont {H.~S.}\ \bibnamefont {Rugo}}, \bibinfo {author}
  {\bibfnamefont {J.~W.}\ \bibnamefont {Park}}, \bibinfo {author}
  {\bibfnamefont {D.~A.}\ \bibnamefont {Hammer}}, \bibinfo {author}
  {\bibfnamefont {G.}~\bibnamefont {Giannone}}, \bibinfo {author}
  {\bibfnamefont {C.~R.}\ \bibnamefont {Bertozzi}}, \ and\ \bibinfo {author}
  {\bibfnamefont {V.~M.}\ \bibnamefont {Weaver}},\ }\href {\doibase
  10.1038/nature13535} {\bibfield  {journal} {\bibinfo  {journal} {Nature}\
  }\textbf {\bibinfo {volume} {511}},\ \bibinfo {pages} {319} (\bibinfo {year}
  {2014})}\BibitemShut {NoStop}%
\bibitem [{\citenamefont {Weikl}\ \emph {et~al.}(2009)\citenamefont {Weikl},
  \citenamefont {Asfaw}, \citenamefont {Krobath}, \citenamefont
  {R\'{o}\.zycki},\ and\ \citenamefont {Lipowsky}}]{Weikl09}%
  \BibitemOpen
  \bibfield  {author} {\bibinfo {author} {\bibfnamefont {T.~R.}\ \bibnamefont
  {Weikl}}, \bibinfo {author} {\bibfnamefont {M.}~\bibnamefont {Asfaw}},
  \bibinfo {author} {\bibfnamefont {H.}~\bibnamefont {Krobath}}, \bibinfo
  {author} {\bibfnamefont {B.}~\bibnamefont {R\'{o}\.zycki}}, \ and\ \bibinfo
  {author} {\bibfnamefont {R.}~\bibnamefont {Lipowsky}},\ }\href@noop {}
  {\bibfield  {journal} {\bibinfo  {journal} {Soft Matter}\ }\textbf {\bibinfo
  {volume} {5}},\ \bibinfo {pages} {3213} (\bibinfo {year} {2009})}\BibitemShut
  {NoStop}%
\bibitem [{\citenamefont {R\'{o}\.{z}ycki}, \citenamefont {Lipowsky},\ and\
  \citenamefont {Weikl}(2010)}]{Rozycki10}%
  \BibitemOpen
  \bibfield  {author} {\bibinfo {author} {\bibfnamefont {B.}~\bibnamefont
  {R\'{o}\.{z}ycki}}, \bibinfo {author} {\bibfnamefont {R.}~\bibnamefont
  {Lipowsky}}, \ and\ \bibinfo {author} {\bibfnamefont {T.~R.}\ \bibnamefont
  {Weikl}},\ }\href@noop {} {\bibfield  {journal} {\bibinfo  {journal} {New J.
  Phys.}\ }\textbf {\bibinfo {volume} {12}},\ \bibinfo {pages} {095003}
  (\bibinfo {year} {2010})}\BibitemShut {NoStop}%
\bibitem [{\citenamefont {Paszek}\ \emph {et~al.}(2009)\citenamefont {Paszek},
  \citenamefont {Boettiger}, \citenamefont {Weaver},\ and\ \citenamefont
  {Hammer}}]{Paszek09}%
  \BibitemOpen
  \bibfield  {author} {\bibinfo {author} {\bibfnamefont {M.~J.}\ \bibnamefont
  {Paszek}}, \bibinfo {author} {\bibfnamefont {D.}~\bibnamefont {Boettiger}},
  \bibinfo {author} {\bibfnamefont {V.~M.}\ \bibnamefont {Weaver}}, \ and\
  \bibinfo {author} {\bibfnamefont {D.~A.}\ \bibnamefont {Hammer}},\ }\href
  {\doibase 10.1371/journal.pcbi.1000604} {\bibfield  {journal} {\bibinfo
  {journal} {PLoS Comput. Biol.}\ }\textbf {\bibinfo {volume} {5}},\ \bibinfo
  {pages} {e1000604} (\bibinfo {year} {2009})}\BibitemShut {NoStop}%
\bibitem [{\citenamefont {Chan}\ and\ \citenamefont {Springer}(1992)}]{Chan92}%
  \BibitemOpen
  \bibfield  {author} {\bibinfo {author} {\bibfnamefont {P.~Y.}\ \bibnamefont
  {Chan}}\ and\ \bibinfo {author} {\bibfnamefont {T.~A.}\ \bibnamefont
  {Springer}},\ }\href@noop {} {\bibfield  {journal} {\bibinfo  {journal} {Mol.
  Biol. Cell}\ }\textbf {\bibinfo {volume} {3}},\ \bibinfo {pages} {157}
  (\bibinfo {year} {1992})}\BibitemShut {NoStop}%
\bibitem [{\citenamefont {Patel}, \citenamefont {Nollert},\ and\ \citenamefont
  {McEver}(1995)}]{Patel95}%
  \BibitemOpen
  \bibfield  {author} {\bibinfo {author} {\bibfnamefont {K.~D.}\ \bibnamefont
  {Patel}}, \bibinfo {author} {\bibfnamefont {M.~U.}\ \bibnamefont {Nollert}},
  \ and\ \bibinfo {author} {\bibfnamefont {R.~P.}\ \bibnamefont {McEver}},\
  }\href@noop {} {\bibfield  {journal} {\bibinfo  {journal} {J. Cell. Biol.}\
  }\textbf {\bibinfo {volume} {131}},\ \bibinfo {pages} {1893} (\bibinfo {year}
  {1995})}\BibitemShut {NoStop}%
\bibitem [{\citenamefont {Huang}\ \emph {et~al.}(2004)\citenamefont {Huang},
  \citenamefont {Chen}, \citenamefont {Chesla}, \citenamefont {Yago},
  \citenamefont {Mehta}, \citenamefont {McEver}, \citenamefont {Zhu},\ and\
  \citenamefont {Long}}]{Huang04}%
  \BibitemOpen
  \bibfield  {author} {\bibinfo {author} {\bibfnamefont {J.}~\bibnamefont
  {Huang}}, \bibinfo {author} {\bibfnamefont {J.}~\bibnamefont {Chen}},
  \bibinfo {author} {\bibfnamefont {S.~E.}\ \bibnamefont {Chesla}}, \bibinfo
  {author} {\bibfnamefont {T.}~\bibnamefont {Yago}}, \bibinfo {author}
  {\bibfnamefont {P.}~\bibnamefont {Mehta}}, \bibinfo {author} {\bibfnamefont
  {R.~P.}\ \bibnamefont {McEver}}, \bibinfo {author} {\bibfnamefont
  {C.}~\bibnamefont {Zhu}}, \ and\ \bibinfo {author} {\bibfnamefont
  {M.}~\bibnamefont {Long}},\ }\href {\doibase 10.1074/jbc.M407039200}
  {\bibfield  {journal} {\bibinfo  {journal} {J. Biol. Chem.}\ }\textbf
  {\bibinfo {volume} {279}},\ \bibinfo {pages} {44915} (\bibinfo {year}
  {2004})}\BibitemShut {NoStop}%
\bibitem [{\citenamefont {Milstein}\ \emph {et~al.}(2008)\citenamefont
  {Milstein}, \citenamefont {Tseng}, \citenamefont {Starr}, \citenamefont
  {Llodra}, \citenamefont {Nans}, \citenamefont {Liu}, \citenamefont {Wild},
  \citenamefont {van~der Merwe}, \citenamefont {Stokes}, \citenamefont
  {Reisner},\ and\ \citenamefont {Dustin}}]{Milstein08}%
  \BibitemOpen
  \bibfield  {author} {\bibinfo {author} {\bibfnamefont {O.}~\bibnamefont
  {Milstein}}, \bibinfo {author} {\bibfnamefont {S.-Y.}\ \bibnamefont {Tseng}},
  \bibinfo {author} {\bibfnamefont {T.}~\bibnamefont {Starr}}, \bibinfo
  {author} {\bibfnamefont {J.}~\bibnamefont {Llodra}}, \bibinfo {author}
  {\bibfnamefont {A.}~\bibnamefont {Nans}}, \bibinfo {author} {\bibfnamefont
  {M.}~\bibnamefont {Liu}}, \bibinfo {author} {\bibfnamefont {M.~K.}\
  \bibnamefont {Wild}}, \bibinfo {author} {\bibfnamefont {P.~A.}\ \bibnamefont
  {van~der Merwe}}, \bibinfo {author} {\bibfnamefont {D.~L.}\ \bibnamefont
  {Stokes}}, \bibinfo {author} {\bibfnamefont {Y.}~\bibnamefont {Reisner}}, \
  and\ \bibinfo {author} {\bibfnamefont {M.~L.}\ \bibnamefont {Dustin}},\
  }\href {\doibase 10.1074/jbc.M804756200} {\bibfield  {journal} {\bibinfo
  {journal} {J Biol Chem}\ }\textbf {\bibinfo {volume} {283}},\ \bibinfo
  {pages} {34414} (\bibinfo {year} {2008})}\BibitemShut {NoStop}%
\bibitem [{\citenamefont {Reister-Gottfried}, \citenamefont {Leitenberger},\
  and\ \citenamefont {Seifert}(2007)}]{Reister07}%
  \BibitemOpen
  \bibfield  {author} {\bibinfo {author} {\bibfnamefont {E.}~\bibnamefont
  {Reister-Gottfried}}, \bibinfo {author} {\bibfnamefont {S.~M.}\ \bibnamefont
  {Leitenberger}}, \ and\ \bibinfo {author} {\bibfnamefont {U.}~\bibnamefont
  {Seifert}},\ }\href@noop {} {\bibfield  {journal} {\bibinfo  {journal} {Phys.
  Rev. E}\ }\textbf {\bibinfo {volume} {75}},\ \bibinfo {pages} {011908}
  (\bibinfo {year} {2007})}\BibitemShut {NoStop}%
\bibitem [{\citenamefont {Naji}, \citenamefont {Atzberger},\ and\ \citenamefont
  {Brown}(2009)}]{Naji09}%
  \BibitemOpen
  \bibfield  {author} {\bibinfo {author} {\bibfnamefont {A.}~\bibnamefont
  {Naji}}, \bibinfo {author} {\bibfnamefont {P.~J.}\ \bibnamefont {Atzberger}},
  \ and\ \bibinfo {author} {\bibfnamefont {F.~L.~H.}\ \bibnamefont {Brown}},\
  }\href@noop {} {\bibfield  {journal} {\bibinfo  {journal} {Phys. Rev. Lett.}\
  }\textbf {\bibinfo {volume} {102}},\ \bibinfo {pages} {138102} (\bibinfo
  {year} {2009})}\BibitemShut {NoStop}%
\bibitem [{\citenamefont {Lipowsky}(1988)}]{Lipowsky88}%
  \BibitemOpen
  \bibfield  {author} {\bibinfo {author} {\bibfnamefont {R.}~\bibnamefont
  {Lipowsky}},\ }\href@noop {} {\bibfield  {journal} {\bibinfo  {journal}
  {Europhys. Lett.}\ }\textbf {\bibinfo {volume} {7}},\ \bibinfo {pages} {255}
  (\bibinfo {year} {1988})}\BibitemShut {NoStop}%
\end{thebibliography}

%

\end{document}